\documentclass[11pt]{article} 
\usepackage{graphicx}
\usepackage{amsfonts, amssymb, amsmath, amsthm, booktabs, dsfont, float, xr-hyper, hyperref, mathrsfs, euscript, enumitem, bm, pifont, multirow, tabularx, multicol, tikz}
\usepackage[onehalfspacing]{setspace}
\usepackage[labelfont=bf,textfont=it]{caption}
\usepackage[labelformat=simple]{subcaption}

\usepackage{lscape}
\hypersetup{pdfborder = {0 0 0},colorlinks=true,allcolors=blue}
\usepackage{natbib}
\usepackage[title]{appendix}
\usepackage[top=1in,bottom=1in,left=1in,right=1in]{geometry}
\usepackage{placeins}
\usepackage{threeparttable}
\usepackage[normalem]{ulem}

\newtheorem{thm}{Theorem}
\newtheorem{coro}{Corollary}
\newtheorem{lem}{Lemma}
\newtheorem{assumption}{Assumption}
\numberwithin{equation}{section}

\newenvironment{examp}[2]
{
    \medskip\par\noindent\textbf{Example #1}\,(#2).
}
{
    \hfill{\LARGE$\lrcorner$}\medskip
}

\theoremstyle{definition}
\newtheorem{remark_tmp}{Remark}
\newenvironment{remark}
	{ \begin{remark_tmp} 	}
	{ 
		\medskip\hfill{\LARGE$\lrcorner$}
		\end{remark_tmp} 
	}
\theoremstyle{definition}
\allowdisplaybreaks[1]

	\DeclareMathOperator*{\argmin}{arg\,min}
	
	\DeclareMathOperator*{\diag}{diag}

\DeclareFontFamily{U}{mathx}{\hyphenchar\font45}
\DeclareFontShape{U}{mathx}{m}{n}{
      <5> <6> <7> <8> <9> <10>
      <10.95> <12> <14.4> <17.28> <20.74> <24.88>
      mathx10
      }{}
\DeclareSymbolFont{mathx}{U}{mathx}{m}{n}
\DeclareFontSubstitution{U}{mathx}{m}{n}
\DeclareMathSymbol{\bigtimes}{1}{mathx}{"91}
	
\renewcommand{\P}{\mathbb{P}}
\newcommand{\E}{\mathbb{E}}
\newcommand{\V}{\mathbb{V}}

\newcommand{\I}{\mathds{1}}

\newcommand{\bA}{\mathbf{A}}
\newcommand{\bB}{\mathbf{B}}
\newcommand{\bC}{\mathbf{C}}

\newcommand{\bG}{\mathbf{G}}

\newcommand{\bI}{\mathbf{I}}

\newcommand{\bQ}{\mathbf{Q}}

\newcommand{\bU}{\mathbf{U}}

\newcommand{\bV}{\mathbf{V}}

\newcommand{\bY}{\mathbf{Y}}
\newcommand{\bZ}{\mathbf{Z}}

\newcommand{\bb}{\mathbf{b}}
\newcommand{\bc}{\mathbf{c}}
\newcommand{\bg}{\mathbf{g}}

\newcommand{\bp}{\mathbf{p}}

\newcommand{\br}{\mathbf{r}}

\newcommand{\bu}{\mathbf{u}}
\newcommand{\bv}{\mathbf{v}}
\newcommand{\bx}{\mathbf{x}}

\newcommand{\bw}{\mathbf{w}}

\newcommand{\ttb}{\star}

\newcommand{\bbeta}{\boldsymbol{\beta}}

\newcommand{\bgamma}{\boldsymbol{\gamma}}

\newcommand{\bSigma}{\boldsymbol{\Sigma}}

\newcommand{\blambda}{\boldsymbol{\lambda}}
\newcommand{\bdelta}{\boldsymbol{\delta}}

\newcommand{\tin}{\mathtt{in}}
\newcommand{\tout}{\mathtt{out}}

\newcommand{\myuline}[1]{\underline{#1\mkern-4mu}\mkern4mu }
\newcommand{\myoline}[1]{\overline{\mkern-3mu#1\mkern-1.5mu}}

\usepackage{algorithm}
\usepackage{algpseudocode}
\newcommand\Algphase[1]{%
\vspace*{-.2\baselineskip}\Statex\hspace*{\dimexpr-\algorithmicindent--4pt\relax}\rule{\textwidth}{0.4pt}%
\Statex\hspace*{\dimexpr-\algorithmicindent--5pt\relax}\textbf{#1}%
\vspace*{-.7\baselineskip}\Statex\hspace*{\dimexpr-\algorithmicindent--4pt\relax}\rule{\textwidth}{0.4pt}%
}

\newcounter{algsubstate}
\renewcommand{\thealgsubstate}{\alph{algsubstate}}
\newenvironment{algsubstates}
  {\setcounter{algsubstate}{0}%
   \renewcommand{\State}{%
     \stepcounter{algsubstate}%
     \Statex\hspace*{\dimexpr-\algorithmicindent--24pt\relax} {\footnotesize\thealgsubstate:}\space}}
  {}

\begin{document}

\title{Uncertainty Quantification in Synthetic Controls\\ with Staggered Treatment Adoption\thanks{We thank the co-Editor, Peter Hull, and three reviewers for their comments. We also thank Alberto Abadie, Simon Freyaldenhoven, and Bartolomeo Stellato for many insightful discussions. Cattaneo and Titiunik gratefully acknowledge financial support from the National Science Foundation (SES-2019432 and SES-2241575), Cattaneo gratefully acknowledges financial support from the National Institute of Health (R01 GM072611-16), and Feng gratefully acknowledges the financial support from the National Natural Science Foundation of China (NSFC) under grants 72203122, 72133002, and 72250064.}\bigskip}
\author{Matias D. Cattaneo\thanks{Department of Operations Research and Financial Engineering, Princeton University.} \and
	    Yingjie Feng\thanks{School of Economics and Management, Tsinghua University.} \and
	    Filippo Palomba\thanks{Department of Economics, Princeton University.} \and
	    Roc{\'i}o Titiunik\thanks{Department of Politics, Princeton University.}}
\maketitle

\begin{abstract}
    We propose principled prediction intervals to quantify the uncertainty of a large class of synthetic control predictions (or estimators) in settings with staggered treatment adoption, offering precise non-asymptotic coverage probability guarantees. From a methodological perspective, we provide a detailed discussion of different causal quantities to be predicted, which we call \textit{causal predictands}, allowing for multiple treated units with treatment adoption at possibly different points in time. From a theoretical perspective, our uncertainty quantification methods improve on prior literature by (i) covering a large class of causal predictands in staggered adoption settings, (ii) allowing for synthetic control methods with possibly nonlinear constraints, (iii) proposing scalable robust conic optimization methods and principled data-driven tuning parameter selection, and (iv) offering valid uniform inference across post-treatment periods. We illustrate our methodology with an empirical application studying the effects of economic liberalization on real GDP per capita for Sub-Saharan African countries. Companion software packages are provided in \texttt{Python}, \texttt{R}, and \texttt{Stata}.
\end{abstract}


\thispagestyle{empty}
\clearpage

\doublespacing
\setcounter{page}{1}
\pagestyle{plain}

\pagebreak
\setcounter{page}{0}\thispagestyle{empty}
\tableofcontents
\pagebreak

\pagestyle{plain}
\section{Introduction} 

The synthetic control (SC) method was introduced by \citet{Abadie-Gardeazabal_2003_AER}, and since then many extensions and generalizations have been proposed \citep[see][and references therein]{Abadie_2021_JEL}. The method is now part of the standard toolkit for program evaluation and treatment effect analysis \citep{Abadie-Cattaneo_2018_ARE}, offering a complement to traditional difference-in-differences, event studies, and other panel data approaches for causal inference with longitudinal aggregate data and few treated units. Most of the synthetic control literature concentrates on identification, as well as on prediction or point estimation of treatment effects, under different causal inference frameworks and algorithmic implementations. In contrast, principled uncertainty quantification of synthetic control predictions or estimators in general settings remains mostly unexplored, despite its importance for empirical work.

Following \citet*{Cattaneo-Feng-Titiunik_2021_JASA}, who proposed non-asymptotic prediction intervals for synthetic controls with a single-treated-unit, we employ a causal inference framework where potential outcomes are assumed to be random and develop novel prediction intervals to quantify the uncertainty of a large class of synthetic control predictions (or estimators) in settings with staggered treatment adoption and multiple treated units. Our contributions include establishing precise non-asymptotic coverage probability guarantees for our novel prediction intervals, introducing scalable robust optimization implementations for possibly nonlinear constraints in the synthetic controls construction, developing principled tuning parameter selection, and proposing valid joint inference methods across time. Inferential procedures with non-asymptotic probability guarantees are valuable because synthetic control applications often have small sample sizes, impeding the applicability of asymptotic approximations. Conceptually, the proposed prediction intervals capture two sources of uncertainty: one coming from the construction of the synthetic control weights with pre-treatment data, and the other generated by the irreducible sampling variability introduced by the post-treatment outcomes. Our prediction intervals also take into account potential misspecification errors explicitly and enjoy other robustness properties due to their non-asymptotic, generic construction. For example, our methods allow for nonlinear regularization in the synthetic controls construction, which accommodates L2 penalization, L1/L2-combined penalization, and other variants thereof. These nonlinear penalization schemes are better suited for application where the covariates exhibit collinearity/codependency, as opposed to L1 penalization schemes, which are better suited for covariate shrinkage/selection.

To motivate our methodological work, we begin by introducing an empirical application in Section \ref{sec: empirical example}, and we use this application throughout to illustrate our contributions. Following \cite{Billmeier-Nannicini_2013_RESTAT}, we investigate the effect of economic liberalization in the 1990s on real GDP per capita for Sub-Saharan African countries. This application includes multiple adoption times and multiple treated units. Furthermore, we consider the special case of outcome-only synthetic controls to improve the exposition and carefully account for the fact that the outcome variable in this application is non-stationary and co-integrated. We also discuss more complex empirical analyses in the supplemental appendix.

Our first contribution is methodological in nature due to the complexity added by the staggered treatment adoption setup, which allows for (but does not require) the existence of multiple treated units changing from control to treatment status at possibly different points in time. In Section \ref{sec: basic framework}, we introduce a basic causal inference framework that is motivated by our empirical application and specifically tailored to synthetic control methods with staggered treatment adoption. Using this framework, we define different causal quantities to be predicted in the context of synthetic controls, which we refer to as \textit{causal predictands}, and explain how prediction methods are implemented. 

Section \ref{sec: prediction interval} then discusses uncertainty quantification in the context of our empirical application and basic causal inference framework. Motivated by our empirical application and recent advances in the synthetic controls literature \citep[see][and references therein]{Abadie_2021_JEL}, our analysis focuses exclusively on incorporating in-sample and out-of-sample uncertainty quantification for outcome-only synthetic control methods with non-stationary data and non-linear constraints. In this setting, we present novel prediction intervals with precise non-asymptotic guarantees for synthetic controls with staggered adoption, under easy-to-interpret sufficient conditions. We also discuss scalable, robust conic optimization implementations of our methods \citep{Boyd_2004_BOOK}, data-driven selection of tuning parameters, and valid joint inference across time.

The main empirical results are presented in Section \ref{sec: empirical results}. Our findings indicate that the economic liberalization in the 1990s did not have a positive economic impact on most emerging Sub-Saharan African countries. This finding is in line with prior empirical results \citep{Billmeier-Nannicini_2013_RESTAT}. The Supplemental Appendix provides additional empirical evidence supporting our main findings, including a re-analysis using alternative synthetic control specifications, placebo treatment dates, donor pool constructions, and other related implementations under different assumptions.

While the basic framework developed in Sections \ref{sec: basic framework} and \ref{sec: prediction interval} is sufficient for our empirical application, Section \ref{sec: theory} offers a substantive generalization of our methods that allows, among others, general misspecification error, multiple covariate features, and cross-equation re-weighting when constructing the synthetic control weights with pre-treatment data. This general framework also provides foundational results for our empirical work, including a theoretical justification for the methods presented in Sections \ref{sec: basic framework} and \ref{sec: prediction interval}.

\subsection{Related Literature}

We contribute to developing prediction/estimation and inference methods for synthetic control settings with staggered treatment adoption. Putting aside generic linear factor model or matrix completion methods, \citet*{BenMichael-Feller-Rothstein_2021_JRSSB}, \citet{Powell_2022_JBES}, and \citet{Shaikh-Toulis_2021_JASA} appear to be the only prior papers that have studied staggered treatment adoption for synthetic controls explicitly. The first paper focuses on prediction/estimation in settings where the pre-treatment fit is poor and develops penalization methods to improve the performance of the canonical synthetic control method. \citet*{BenMichael-Feller-Rothstein_2021_JRSSB} also suggest employing a bootstrap method for assessing uncertainty, but no formalization is provided guaranteeing its (asymptotic) validity. \citet{Powell_2022_JBES} employs a standard parametric panel data model to discuss large-sample estimation and inference methods for a single common treatment effect. \cite{Shaikh-Toulis_2021_JASA} focus on uncertainty quantification employing a parametric duration model and propose a permutation-based inferential method under a symmetry assumption. Our paper complements these prior contributions by (i) developing a general causal inference framework for synthetic control methods with staggered treatment adoption, and (ii) offering nonparametric inference methods with demonstrable non-asymptotic coverage guarantees and allowing for misspecification in the construction of the synthetic control weights. We also propose novel scalable robust conic optimization implementations, principled tuning parameter selection methods, and a valid joint inference procedure across multiple time periods.

Our quantification of uncertainty via non-asymptotic prediction intervals follows \citet*{Cattaneo-Feng-Titiunik_2021_JASA}. These inference methods are motivated by \citet{Vovk_2012_ACML} and are most closely related to prior work by \citet*{Chernozhukov-Wuthrich-Zhu_2021_PNAS} and \citet*{Chernozhukov-Wuthrich-Zhu_2021_JASA} on conformal prediction intervals. Relative to this literature, our contributions include: (i) allowing for a large class of causal predictands in staggered adoption settings (prior work covered only the canonical single treated unit case); (ii) covering a large class of synthetic control predictions with possibly nonlinear constraints (prior work allowed only linear constraints); (iii) developing scalable robust optimization implementations and proposing principled data-driven tuning parameter selection (prior work did not provide guidance on these issues); and (iv) introducing valid uniform inference methods across post-treatment periods (absent in prior work).

There are a few other, conceptually different, recent proposals to quantify uncertainty and conduct inference in the synthetic controls literature. For example, \cite{Li_2020_JASA} study correctly specified linear factor models, \citet{Masini-Medeiros_2021_JASA} study high-dimensional penalization methods, \citet*{Agarwal-Shah-Shen-Song_2021_JASA} investigate matrix completion methods, and \citet*{shi2023theory} develop inference methods using a proximal causal inference framework. All these methods rely on asymptotic approximations, in most cases employing standard Gaussian critical values that assume away misspecification errors and other small sample issues. Our work complements these contributions by providing prediction intervals with non-asymptotic coverage guarantees \citep{Wainwright_2019_Book}. There is also a literature in econometrics on event studies that is loosely related to synthetic controls with staggered adoption: see, for example, \cite{FreyaldenhovenEtal2019-AER,FreyaldenhovenEtal2024-wp}, \citet{Miller2023-JEP}, and references therein. Finally, all the inferential methods mentioned so far contrast with the original method proposed by \citet*{Abadie-Diamond-Hainmueller_2010_JASA}, which relies on design-based permutation of treatment assignment assuming that the potential outcomes are non-random.

\subsection{Paper Organization}

Section \ref{sec: empirical example} introduces our running empirical application. Section \ref{sec: basic framework} presents the basic causal inference framework for outcome-only synthetic controls with staggered treatment adoption, and Section \ref{sec: prediction interval} discusses non-asymptotic uncertainty quantification in that context. Section \ref{sec: empirical results} presents our empirical results. Section \ref{sec: theory} gives a self-contained discussion of our most general framework and theoretical results. The Supplemental Appendix includes all proofs, additional empirical results, and other technical details omitted to improve the exposition.

We provide general-purpose software implementing our results in \texttt{Python}, \texttt{R}, and \texttt{Stata}, including detailed documentation and additional replication materials. This software is discussed in our companion article \citet*{Cattaneo-Feng-Palomba-Titiunik_2025_JSS}, where we addressed several implementation issues related to numerical optimization and tuning parameter selection. To complement the illustration, in Section S.7 of the Supplemental Appendix, we provide more details on how to prepare the data to analyze staggered treatment adoption using synthetic control methods using our companion software. In addition, in the Supplemental Appendix, Section S.2.3 demonstrates how to obtain tighter bounds for uncertainty quantification when the weighting matrix has a block diagonal structure, and Section S.6 shows how to reformulate the synthetic control problem as a scalable conic optimization problem to improve computational efficiency. For ease of understanding, Table S.1 in the Supplemental Appendix summarizes the notation used throughout the paper.

\section{The Effect of Liberalization on GDP for Sub-Saharan African Countries}
\label{sec: empirical example}

During the second half of the twentieth century, many countries around the world launched programs of (external) economic liberalization, booming from 22\% in 1960 to 73\% in the early 2000s \citep{Wacziarg-Welch_2008_WBER}. In the last thirty years, scholars have investigated the social and economic consequences of such liberalization programs, often reaching conflicting conclusions \citep*[see, e.g., ][]{Levine-Renelt_1992_AER, Sachs-Warner_1995_BP, Dejong-Ripoll_2006_RES}.

The impact of liberalization policies on economic welfare has been traditionally investigated with cross-country analyses \citep*[e.g.][]{Sachs-Warner_1995_BP} and individual case studies \citep[e.g.][]{Bhagwati-Srinivasan_2001_book}. More recently, scholars have turned to synthetic control methods in the hope of employing a causal inference methodology that allows for the presence of time-varying unobservable confounders. Employing the synthetic control framework originally developed in \cite{Abadie-Gardeazabal_2003_AER}, \cite{Billmeier-Nannicini_2013_RESTAT} analyzed the effects of liberalization in four continents: Africa, Asia, North America, and South America. They used a pre-existing dataset of economic variables \citep[previously used in][]{Giavazzi-Tabellini_2005_JME} which includes 180 countries, covers the period 1963--2000 and contains an indicator for economic liberalization originally defined in \citet*{Sachs-Warner_1995_BP} and updated in \cite{Wacziarg-Welch_2008_WBER} (hereafter, the Sachs-Warner indicator). More details on the data and the definition of economic liberalization can be found in Section S.7 of the Supplemental Appendix.

\cite{Billmeier-Nannicini_2013_RESTAT} studied the effect of economic liberalization---as measured by the Sachs-Warner indicator---on GDP per capita, to understand whether the adoption of liberalization programs affected economic growth. Building on their study of Sub-Saharan Africa, we illustrate how our formal inference framework can be an essential tool to aggregate results and draw general conclusions from synthetic control analyses under staggered treatment adoption by multiple treated units. The ambitious scope of the analysis in \cite{Billmeier-Nannicini_2013_RESTAT} resulted in a large number of synthetic control results. In Sub-Saharan Africa alone, they studied sixteen episodes of liberalization that occurred at ten different periods. The results exhibited considerable heterogeneity, which the authors summarized by grouping the effects into four categories according to two dimensions: whether the effect was positive and whether there was evidence that the effect was robust rather than ``coincidental'' (a notion of statistical significance). The four categories were (a) countries with a positive and strongly robust effect, (b) countries with a positive and somewhat robust effect, (c) countries with a positive but non-robust effect, and (d) countries with a null effect. They concluded that ``Botswana is the only country clearly in the first group, offering a truly convincing success story'' (p. 995).

\cite{Billmeier-Nannicini_2013_RESTAT} adjudicated the robustness or statistical significance of each of the sixteen effects with the Fisherian-type placebo test developed by \cite{Abadie-Diamond-Hainmueller_2010_JASA}, which was the main inference tool available at the time. Our new inference framework allows us to reconsider the evidence for Sub-Saharan Africa by providing formal tools to perform joint inference and draw a general conclusion about the effects of liberalization by considering the evidence altogether rather than piecemeal.  

The data starts in the 1960s, which marks the beginning of Africa's post-colonial era. Since independence from their colonial rulers until the late 1980s, many countries in Sub-Saharan Africa adopted neo-patrimonial political regimes based on the concentration of power in a single individual and the resulting cultivation of personalistic politics, widespread clientelistic networks, and the use of the resources of the state to achieve political legitimacy \citep{Bratton1VandeWalle1997-book}. 

The 1980s put high stress on these regimes, as economic conditions deteriorated. Negative economic growth, increased debt services as a proportion of exports, currency overvaluation, low commodity prices, low investment, and limited access to credit created an adverse economic environment that these non-democratic regimes were ill-equipped to handle. International finance organizations such as the World Bank and the International Monetary Fund incentivized programs of structural economic adjustment, with disputed success \citep{CallaghyRavenhill1994-book}.

The adverse economic conditions led to popular protests in the early 1990s, which marked the beginning of a wave of transitions to democracy throughout the continent. Starting in the early 1990s, many countries started a process of political liberalization, partly influenced by the fall of the Berlin Wall in November of 1989 and the subsequent collapse of the communist regimes of the Soviet Union \citep{Bratton1VandeWalle1997-book}.

As illustrated in Figure \ref{fig: staggeredAfrica}, most countries in Sub-Saharan Africa adopted the economic liberalization treatment between the late 1980s and the early or mid-1990s which, as just explained, was a period of major political and economic transition in Africa and beyond. This means that, for many countries in our sample, the economic liberalization treatment occurred nearly simultaneously with other major political and economic shocks that influenced the World's geo-political equilibrium. Given these potential confounders, it is crucial to exploit the staggered adoption of the treatment to make inferences. While the worldwide political changes between the late 1980s and early 1990s affected all countries at the same time, the concrete measures of economic liberalization captured by the Sachs index were introduced in different years for different countries. Our proposed synthetic control methods allow us to define causal predictands and perform joint inferences that leverage the staggered introduction of the treatment, while controlling for worldwide shocks affecting all units.

\begin{figure}[!ht]
    \caption{Staggered Treatment Adoption in Africa.}
    \label{fig: staggeredAfrica}
        \resizebox{0.9\textwidth}{!}{%
        \input{figs/africa_matrix_treatment.tex}
        }
        \par
 	\begin{center}
 		\parbox[1]{\textwidth}{\footnotesize \textit{Notes:} A country $i$ in year $t$ is defined to be ``closed" or ``open" using to the Sachs index. Thick vertical lines--corresponding to 1987 and 1991--delimit the three main liberalization waves of \cite{Billmeier-Nannicini_2013_RESTAT}.}
 	\end{center}  
 \end{figure}

\section{Basic Framework}\label{sec: basic framework}

We set up a basic synthetic control framework that matches the empirical application in Section \ref{sec: empirical example} and suffices to communicate key ideas of our proposed method. A more general and technically more involved framework is deferred to Section \ref{sec: theory}.

Suppose that we observe $N$ countries (``units'') for $T$ years (``periods''). Countries are indexed by $i=1, \ldots, N$, and years are indexed by $t=1, \ldots, T$. For each country $i$ in year $t$, we can observe the GDP per capita $Y_{it}$ (``outcome'') and a time $T_i$ that indicates when country $i$ adopted economic liberalization (``treatment'').
Assume that a country remains economically liberalized whenever $t\geq T_i$. (In our empirical application this is always verified.)
Without loss of generality, countries are ordered in the adoption times so that $1\leq T_1\leq T_2\leq \cdots \leq T_N\leq \infty$, with $T_i=\infty$ denoting that country $i$ remains untreated throughout the observation period.
Let $\mathcal{N}=\{i: T_i=\infty\}$ be the group of ``never-liberalized'' countries and 
$\mathcal{E}=\{i: T_i<\infty\}$ the group of ``ever-liberalized'' countries. 
Define $J_0=|\mathcal{N}|$ and $J_1=|\mathcal{E}|$, where we use $|\mathcal{A}|$ to denote the number of elements in $\mathcal{A}$ for any finite set $\mathcal{A}$.

We are interested in the effect of economic liberalization on a country's GDP per capita. Let $Y_{it}(s)$ denote the potential GDP per capita of country $i$ in year $t$ that would be observed had country $i$ adopted the economic liberalization treatment in year $s$, for $s=1, \ldots, T, \infty$, and assume $Y_{it}(s)=Y_{it}(\infty)$ whenever $t<s$. Implicitly, these simplifications impose two standard assumptions: no spillovers (the potential outcomes of country $i$ depend only on $i$’s adoption time) and no anticipation (a country's potential outcomes prior to the treatment are equal to the outcomes it would have had if it had never been treated). The observed GDP per capita $Y_{it}$ can be written as
\[
Y_{it}=Y_{it}(\infty)\I(t<T_i)+Y_{it}(T_i)\I(t\geq T_i).
\]
That is, whenever the economic liberalization has not been adopted, we always observe the potential GDP per capita under the ``never-treated'' status; otherwise the potential GDP per capita corresponding to the adoption time $T_i$ is observed.

Our primary goal is to make a prediction of the missing counterfactual GDP per capita $Y_{it}(\infty)$ for an ever-liberalized country in the post-treatment period; we then use this unit-level predictand as the basis of a variety of other aggregate predictands.

For simplicity, we take all never-treated units in $\mathcal{N}$ as ``donors'' for SC prediction, 
and let all years prior to the economic liberalization, 
$\{t: 1\leq t \leq T_i-1\}$, be the training period used to obtain the SC weights for each ever-liberalized country $i\in\mathcal{E}$.
The idea of SC is to find a vector of weights on the donor countries such that the weighted average of pre-treatment GDP per capita of donor countries matches that of the treated country as closely as possible, subject to some regularization constraints. Let $\|\cdot\|_1$ and $\|\cdot\|_2$ denote the usual L1 and L2 vector norms respectively,  and $(v_j: j\in\mathcal{A})$ denote a vector consisting of all $v_j$'s with $j\in\mathcal{A}$ for a set $\mathcal{A}$.  
In our empirical analysis below, the SC weights are obtained via an L1-L2 constrained regression:
\begin{equation}\label{eq: weights in basic framework}
\widehat{\bw}^{[i]}=\underset{\bw\in\mathcal{W}^{[i]}}{\arg\min}\;
\sum_{t=1}^{T_i-1}
\Big(Y_{it}-\bY_{\mathcal{N}t}'\bw\Big)^2
\quad \text{for each }\; i\in\mathcal{E},
\end{equation}
where $\bY_{\mathcal{N}t}=(Y_{jt}: j\in\mathcal{N})$ is the vector of GDP per capita for donor countries at time $t$, and 
$\mathcal{W}^{[i]}=\{\bw\in\mathbb{R}^{J_0}_+: \|\bw\|_1= 1, \|\bw\|_2\leq Q^{[i]}\}$ for some constant $Q^{[i]}>0$. The choice $Q^{[i]}$ is discussed in Section S.7.2 of the Supplemental Appendix. The prediction of country $i$'s counterfactual GDP per capita $k$ periods after the economic liberalization is given by
\[
\widehat{Y}_{i(T_i+k)}(\infty)=\bY_{\mathcal{N}(T_i+k)}'\widehat{\bw}^{[i]},\quad 
\text{for each}\;\; i\in\mathcal{E} \;\;\text{and}\;\; 0\leq k\leq T-T_i.
\]

In what follows, we define a variety of causal predictands and construct their SC predictions accordingly.

\begin{examp}{TSUS}{Time-specific unit-specific predictand}
\hypertarget{exmp:tsus}{}
    The first predictand we consider is the effect of the intervention for a specific unit in a specific time period, which is the primary causal predictand of interest in classical SC analysis with only one treated unit.

    For each ever-treated unit $i\in\mathcal{E}$, we can define the TSUS predictand in each post-treatment period:
    \[
    \tau_{ik}:=Y_{i(T_i+k)}(T_i)-Y_{i(T_i+k)}(\infty), \quad 0\leq k\leq T-T_i.
    \]
    In our empirical application, this predictand captures the effect of economic liberalization on a specific country, measured $k$ years after the adoption of the liberalization policy.  
    Given the prediction $\widehat{Y}_{i(T_i+k)}(\infty)$, we predict $\tau_{ik}$ by
    $$\widehat{\tau}_{ik}:=Y_{i(T_i+k)}(T_i)-\widehat{Y}_{i(T_i+k)}(\infty).$$
    See Figure \ref{fig: predictand-TSUS} for a graphical representation of $\widehat{\tau}_{ik}$.
\end{examp}

\begin{examp}{TAUS}{Time-averaged unit-specific predictand}
\hypertarget{exmp:taus}{} 
    When there are multiple post-treatment periods, scholars may be interested in the effect on a treated unit averaged across all periods.  This defines the TAUS predictand, which takes the average over time of the TSUS predictands for an ever-treated unit $i$:
    $$\tau_{i\cdot}:=\frac{1}{T-T_i+1}\sum_{k=0}^{T-T_i}\tau_{ik}.$$
    In our empirical application, this is the effect of economic liberalization on a specific country, averaged over the entire post-liberalization period.
    Given the TSUS prediction for country $i$ in period $k$, we predict $\tau_{i\cdot}$ by
    $$\widehat{\tau}_{i\cdot}:=\frac{1}{T-T_i+1}\sum_{k=0}^{T-T_i}\widehat{\tau}_{ik}.$$
    See Figure \ref{fig: predictand-TAUS} for a graphical representation of $\widehat{\tau}_{i\cdot}$.
\end{examp}       
 
\begin{examp}{TSUA}{Time-specific unit-averaged predictand}
\hypertarget{exmp:tsua}{}
    When there are multiple treated units, it is also of interest to define a predictand that captures the average effect of the intervention across a given group of units, at a single period in time.
   Let $\mathcal{Q}\subseteq\{1, 2, \cdots, T\}$ be a set of adoption times. The TSUA predictand is defined as
    $$\tau_{\mathcal{Q}k}:=\frac{1}{Q}\sum_{i:T_i\in \mathcal{Q}}\tau_{ik}, \qquad
    Q=|\{i: T_i\in\mathcal{Q}\}|.$$
    In our empirical application, this is the average effect of economic liberalization on the group of countries with adoption times in $\mathcal{Q}$, measured $k$ years after liberalization. In later empirical analysis we consider the treatment effect on countries that liberalized in three different waves---before 1987, between 1987 and 1991, and after 1991.
    Given the TSUS predictions, we predict $\tau_{\mathcal{Q}k}$ by
    $$\widehat{\tau}_{\mathcal{Q}k}:=\frac{1}{Q}\sum_{i: T_i\in\mathcal{Q}}
    \widehat{\tau}_{ik}.$$
    See Figure \ref{fig: predictand-TSUA} for a graphical representation of $\widehat{\tau}_{\mathcal{Q}k}$.
\end{examp}

\begin{examp}{TAUA}{Time-averaged unit-averaged predictand}
    Finally, when there are multiple treated units and multiple post-treatment periods, we may be interested in a predictand that captures the overall average effect of the intervention.

    We define the TAUA predictand as the average of the TSUS predictands across all treated units and over $L$ post-treatment periods:
    $$\tau_{\cdot\cdot}:=\frac{1}{LJ_1}\sum_{k=1}^L\sum_{i\in\mathcal{E}}\tau_{ik}.$$
    In our empirical application, this is the average effect of economic liberalization on all ever-liberalized countries over $L$ years after the policy adoption.
    We assume all ever-treated units are observed at least $L$ periods after the adoption for some $L\geq 0$.
    However, since the observation ends at time $T$, some ever-treated units may have to be excluded in this definition as $L$ varies.      
    Given the TSUS predictions, we predict $\tau_{\cdot\cdot}$ by
    \[\widehat{\tau}_{\cdot\cdot}:=
    \frac{1}{LJ_1}\sum_{k=1}^L\sum_{i\in\mathcal{E}}\widehat{\tau}_{ik}.\]
\end{examp}

We conclude this section with a final note on the nature of the predictands defined above. The potential outcomes, treatment adoption times, and individual effects are all viewed as \textit{random} quantities. We assume that there is only a fixed (possibly small) number of ever-treated units and time periods, which is often the case in synthetic control analysis and accommodates our empirical application. Thus, the various predictands defined above are also random quantities in general, which is why we prefer referring to them as ``predictands'' rather than as treatment ``effects''. However, we do occasionally use the term ``effect'' or ``predicted effect'' to emphasize its randomness and to maintain consistency with the term used for analogous quantities defined in the SC literature under a fixed, non-random potential outcomes framework.  
In classical large-sample causal analysis, target parameters are often probability or ergodic (non-random) limits of the average effects above as $Q\rightarrow\infty$, $J_1\rightarrow\infty$, and/or $T-T_i\rightarrow\infty$. Although our results are also valid in such large-sample settings, we develop statistical inference methods based on prediction intervals that describe a region where a new realization of a random causal predictand of interest is likely to be observed, rather than the usual confidence intervals giving a region in the parameter space for a non-random parameter of interest.

\begin{figure}\label{fig: predictand}
    \caption{Graphical Representation of the Predictands.}
     \centering
     \begin{subfigure}[b]{\textwidth}
         \centering
         \resizebox{0.9\textwidth}{!}{%
         \input{figs/ill_unit_time.tex}}
         \caption{Time-specific unit-specific (TSUS) predictand.}
         \label{fig: predictand-TSUS}
     \end{subfigure}
     \\\bigskip
     \begin{subfigure}[b]{\textwidth}
         \centering
         \resizebox{0.9\textwidth}{!}{%
         \input{figs/ill_unit.tex}}
         \caption{Time-averaged unit-specific (TAUS) predictand.}
         \label{fig: predictand-TAUS}
     \end{subfigure} \\\bigskip
     \begin{subfigure}[b]{\textwidth}
         \centering
         \resizebox{0.9\textwidth}{!}{%
         \input{figs/ill_time.tex}}
         \caption{Time-specific unit-averaged (TSUA) predictand.}
         \label{fig: predictand-TSUA}
     \end{subfigure}
\end{figure}

\section{Uncertainty Quantification}\label{sec: prediction interval}

Let $\tau$ denote any of the four causal predictands defined above. 
Our goal is to construct a (random) prediction interval $\mathcal{I}$ such that, with some high probability $(1-\pi)$ over a conditioning set $\mathscr{H}$, $\mathcal{I}$ covers $\tau$ with a pre-specified probability $(1-\alpha)$ given $\mathscr{H}$: 
\begin{equation}\label{eq: PI definition}
    \P\Big\{\P\big[\tau\in\mathcal{I} \,\big|\, \mathscr{H}\big]\geq 1-\alpha\Big\}\geq 1-\pi.
\end{equation}
Generally, the choice of the conditioning set $\mathscr{H}$ determines the uncertainty that would not be taken into account by the prediction interval.
If we do not condition on anything, technically, $\mathscr{H}$ is a trivial $\sigma$-field, and $\mathcal{I}$ reduces to an unconditional prediction interval.
Since we regard synthetic controls as a regression-based out-of-sample forecasting problem, where it is standard to condition on the ``covariates'' (outcomes of donor countries in this context), we focus on the uncertainty from the ever-treated countries.
Thus, we let $\mathscr{H}=\{Y_{it}: i\in\mathcal{N}, 1\leq t\leq T\}$. 
Also, in practice one needs to set a desired (conditional) coverage level $(1-\alpha)$, say $90\%$, whereas the probability loss $\pi$ over $\mathscr{H}$ is a ``small'' number that only needs to be theoretically characterized. 
In this paper, all results are valid if the training period is long enough ($T_i-1$ is large), with the associated probability $1-\pi$ characterized precisely. 
Thus, we say that the conditional prediction interval offers finite-sample probability guarantees. Our results imply that $\pi\to 0$ as $T_i\to\infty$, but no limits or asymptotic arguments are used in this paper.

To better understand the sources of uncertainty of SC predictions, 
we define the target quantity of the SC weights (conditional on $\mathscr{H}$) that is analogous to \eqref{eq: weights in basic framework}:
\begin{equation}\label{eq: pseudo true value in basic framework}
\bw_0^{[i]} = \underset{\bw\in\mathcal{W}^{[i]}}{\arg\min}\; 
\E\left[\,\sum_{t=1}^{T_i-1}\Big(Y_{it}-\bY_{\mathcal{N}t}'\bw\Big)^2\bigg|\mathscr{H}\right].
\end{equation}
Thus, we can write 
\begin{equation}\label{eq: decomposition of Y}
Y_{it}(\infty)=\bY_{\mathcal{N}t}'\bw_0^{[i]}+u_{it},\quad 
i\in\mathcal{E}, \; 1\leq t\leq T,
\end{equation}
where $u_{it}$ is the corresponding pseudo-true residual relative to the conditioning set $\mathscr{H}$.
Then, for each ever-treated country $i\in\mathcal{E}$,
we have the following decomposition of the counterfactual outcome prediction:
$$\widehat{Y}_{i(T_i+k)}(\infty)-Y_{i(T_i+k)}(\infty)
=\bY_{\mathcal{N}(T_i+k)}'(\widehat{\bw}^{[i]}-\bw_0^{[i]})-u_{i(T_i+k)},
$$
where $\bY_{\mathcal{N}(T_i+k)}'(\widehat{\bw}^{[i]}-\bw_0^{[i]})$ captures the \textit{in-sample uncertainty} from the SC weights construction using the pre-treatment information, and $u_{i(T_i+k)}$ captures the \textit{out-of-sample uncertainty} from the stochastic error in a specific post-treatment period. 
Notice that the in-sample uncertainty quantification is necessary in this scenario since the conditioning set $Y_{it}\not\in\mathscr{H}$ for $i\in\mathcal{E}$.

Accordingly, a similar decomposition can be performed for the prediction $\widehat{\tau}$ of $\tau$:
\[
\widehat{\tau}-\tau=\mathsf{InErr}(\tau)+\mathsf{Out Err}(\tau),
\]
where $\mathtt{InErr}(\tau)$ and $\mathtt{OutErr}(\tau)$ denote the in-sample error and the out-of-sample error associated with the prediction of $\tau$ respectively. Specific expressions of such errors for each treatment effect prediction are given below:
\begin{alignat*}{2}
    &\mathtt{InErr}(\tau_{ik})=
    -\bY_{\mathcal{N}(T_i+k)}'(\widehat{\bw}^{[i]}-\bw_0^{[i]}),
    \qquad
    &&\mathtt{OutErr}(\tau_{ik})=u_{i(T_i+k)},    \\
    &\mathtt{InErr}(\tau_{i\cdot})=
    -\frac{1}{T-T_i+1}\sum_{k=0}^{T-T_i}\bY_{\mathcal{N}(T_i+k)}'(\widehat{\bw}^{[i]}-\bw_0^{[i]}),\qquad
    &&\mathtt{OutErr}(\tau_{i\cdot})=\frac{1}{T-T_i+1}\sum_{k=0}^{T-T_i}u_{i(T_i+k)},\\
    &\mathtt{InErr}(\tau_{\mathcal{Q}k})=
    -\frac{1}{Q}\sum_{i:T_i\in\mathcal{Q}}\bY_{\mathcal{N}(T_i+k)}'(\widehat{\bw}^{[i]}-\bw_0^{[i]}),\qquad
    &&\mathtt{OutErr}(\tau_{\mathcal{Q}k})=\frac{1}{Q}\sum_{i: T_i\in\mathcal{Q}}u_{i(T_i+k)},\\
    &\mathtt{InErr}(\tau_{\cdot\cdot})=
    -\frac{1}{LJ_1}\sum_{k=1}^L\sum_{i\in\mathcal{E}}
    \bY_{\mathcal{N}(T_i+k)}'(\widehat{\bw}^{[i]}-\bw_0^{[i]}),\qquad
    &&\mathtt{OutErr}(\tau_{\cdot\cdot})=\frac{1}{LJ_1}\sum_{k=1}^L\sum_{i\in\mathcal{E}}u_{i(T_i+k)}.
\end{alignat*}

The target SC weights $\bw_0^{[i]}$ and the residual $u_{it}$ need to be understood in context. In principle, the weights $\bw_0^{[i]}$ represent a ``stable'' cross-sectional relationship among treated and donor units that can be learned in the training period and translated to the post-treatment period, which is the common feature of all SC methods. In our empirical application, the outcome $Y_{it}$ is GDP per capita, which is well known to be a non-stationary time series. Then, the idea of SC appears to be more applicable when GDP per capita sequences of different countries are cointegrated, where the ``stable relationship'' is given by the cointegrating vector, and the remainder $u_{it}$ is a stationary series. 
Thus, we make the following assumption on the data generating process:
\begin{assumption}[Data generating process]\label{assumption: dgp}
    Assume that for each $j\in\mathcal{N}$, 
    $Y_{jt}=Y_{j(t-1)}+v_{jt}$, for each $i\in\mathcal{E}$, $Y_{it}=\bY_{\mathcal{N}t}'\bw_0^{[i]}+u_{it}$,
    and $(\bu_t', \bv_t')'$ is i.i.d. over $t$, 
    where $\bu_t=(u_{1t}, \cdots u_{J_1t})'$ and 
    $\bv_t=(v_{(J_1+1)t}, \cdots, v_{Nt})'$. 
    Assume $\bv_t$ is sub-Gaussian, and  
    $u_{it}$ is sub-Gaussian conditional on $\mathscr{H}$ with parameter $\sigma_{it}$ for each $i\in\mathcal{E}$.
\end{assumption}
The outcomes of donor units are assumed to follow the simple unit root process with sub-Gaussian errors; other non-stationary patterns can also be accommodated at the cost of more technicalities. The conditional sub-Gaussianity of $u_{it}$, as precisely defined in Section \ref{sec: out-of-sample error}, enables us to characterize the tail probability of $u_{it}$ exceeding a given bound, which is useful for quantifying the out-of-sample uncertainty, but can also be replaced with weaker moment conditions. We emphasize that Assumption \ref{assumption: dgp} is motivated by our empirical application and only used to illustrate the key idea of our method; our theory is more general and can be applied to other types of data, e.g., weakly dependent time series satisfying certain mixing conditions while allowing for a certain degree of model misspecification. See more discussion in Sections \ref{sec: in-sample error} and \ref{sec: theory} below. Also, we view \eqref{eq: decomposition of Y} as a \textit{predictive} model, and the cointegrated system is merely one way to justify the SC approach; in the Supplemental Appendix Section S.3, we briefly discuss an alternative justification that assumes a linear factor model, and explain how to interpret the in-sample and out-of-sample uncertainty in that scenario.

We employ an intuitive strategy to construct prediction intervals for any causal predictand $\tau$ defined before. 
Specifically, if we can find random intervals 
$[\myuline{M}_{\tin}(\tau), \, \myoline{M}_{\tin}(\tau)]$ and 
$[\myuline{M}_{\tout}(\tau), \, \myoline{M}_{\tout}(\tau)]$ that (conditionally) cover the in-sample error $\mathtt{InErr}(\tau)$ and the out-of-sample error 
$\mathtt{OutErr}(\tau)$, respectively, with certain probabilities, i.e.,
\begin{alignat*}{4}
	&\P\Big\{\P\big[\myuline{M}_{\tin}(\tau)
 &&\leq\;\mathsf{InErr}(\tau)\;
 &&\leq\myoline{M}_{\tin}(\tau) \; &&\big| \; \mathscr{H} \big]\geq 1-\alpha_{\tin}\Big\}\geq 1-\pi_{\tin} \qquad \text{and}\\
	&\P\Big\{\P\big[\myuline{M}_{\tout}(\tau)
 &&\leq \mathsf{OutErr}(\tau)&&\leq \myoline{M}_{\tout}(\tau) \; &&\big| \; \mathscr{H} \big] \geq 1-\alpha_{\tout}\Big\}\geq 1-\pi_{\tout},
	\end{alignat*} 
then, by a union bound we have 
$$
\P\Big\{\P\big[\widehat{\tau}-\myoline{M}_{\tin}(\tau)-\myoline{M}_{\tout}(\tau) \leq \tau\leq \widehat{\tau}-\myuline{M}_{\tin}(\tau)-\myuline{M}_{\tout}(\tau) \big|  \mathscr{H}  \big]
	\geq 1-\alpha_{\tin}-\alpha_{\tout}\Big\}\geq 1-\pi_{\tin}-\pi_{\tout}.$$
That is, the prediction interval 
$\mathcal{I}(\tau):=[\widehat{\tau}-\myoline{M}_{\tin}(\tau)-\myoline{M}_{\tout}(\tau),\;
\widehat{\tau}-\myuline{M}_{\tin}(\tau)-\myuline{M}_{\tout}(\tau)]$ achieves $(1-\alpha_{\tin}-\alpha_{\tout})$ coverage probability conditional on $\mathscr{H}$, 
which holds with probability at least 
$(1-\pi_{\tin}-\pi_{\tout})$ over $\mathscr{H}$.
In practice, we can set, for example, 
$\alpha_{\tin}=\alpha_{\tout}=\alpha/2$ to achieve the desired coverage level 
$1-\alpha$, and our theory will precisely characterize $\pi_{\tin}+\pi_{\tout}$ and ensure that it is small at least when the training period is long enough. 
Bounding two errors separately makes the resultant prediction interval for $\tau$ conservative, but it clearly differentiates the contribution of the SC training procedure and the out-of-sample prediction procedure to the final inference.

\subsection{In-Sample Uncertainty Quantification}\label{sec: in-sample error}
We begin with the in-sample uncertainty quantification for the SC prediction of the time-specific counterfactual outcome $Y_{i(T_i+k)}(\infty)$ for each ever-treated unit, and then extend it to the four causal predictands defined before.
Importantly, we take the decomposition 
\eqref{eq: decomposition of Y} as a predictive model with no structural meaning, which does not have to be a ``correctly specified'' mean regression model. That is, we allow the outcomes of the donor countries $\bY_{\mathcal{N}t}$ to be possibly correlated with the SC residual $u_{it}$, i.e., $\bgamma^{[i]}:=\E[\bY_{\mathcal{N}t}u_{it}|\mathscr{H}]\neq \mathbf{0}$, which violates the standard identification assumption for mean regression models. Accommodating this possibility is important, especially in high-dimensional scenarios, since SC constraints in $\mathcal{W}^{[i]}$ are usually imposed to make SC predictions feasible, and probably more stable, rather than being grounded in prior knowledge of the true data generating process. 

Our proposed uncertainty quantification strategy only makes use of three basic facts: 
(i) $\widehat{\bw}^{[i]}$ is the optimizer of the sample-based SC problem \eqref{eq: weights in basic framework};
(ii) $\bw_0^{[i]}$ is the optimizer of the population-based SC problem \eqref{eq: pseudo true value in basic framework}; and (iii) the L1-L2 constraint set $\mathcal{W}^{[i]}$ used in our empirical application is convex. 
Actually, (i) and (ii) are true by definition of SC, and (iii) are true for all commonly used constraints, such as simplex, L1 constraint, L2 constraint, or some combinations thereof. 
Consequently, it can be shown that the following inequality holds (deterministically): 
$$(\widehat{\bw}^{[i]}-\bw_0^{[i]})'\widehat{\bQ}^{[i]}(\widehat{\bw}^{[i]}-\bw_0^{[i]})-
2(\widehat{\bgamma}^{[i]}-\bgamma^{[i]})'(\widehat{\bw}^{[i]}-\bw_0^{[i]})\leq0\quad \text{for each }i\in\mathcal{E},$$ 
where $\widehat{\bQ}^{[i]}=\sum_{t=1}^{T_i-1}\bY_{\mathcal{N}t}\bY_{\mathcal{N}t}'$ and
$\widehat\bgamma^{[i]}=\sum_{t=1}^{T_i-1}\bY_{\mathcal{N}t}u_{it}$.
Therefore, the following bounds on the in-sample error of the prediction  $\widehat{Y}_{i(T_i+k)}(\infty)$ hold:
$$
\inf_{\bdelta\in\mathcal{M}_{\widehat{\bgamma}}^{[i]}}\,\bY_{\mathcal{N}(T_i+k)}'\bdelta
\;\leq \bY_{\mathcal{N}(T_i+k)}'(\widehat{\bw}^{[i]}-\bw_0^{[i]})\;\leq 
\sup_{\bdelta\in\mathcal{M}_{\widehat{\bgamma}}^{[i]}}\,\bY_{\mathcal{N}(T_i+k)}'\bdelta,
$$
where
$\mathcal{M}_{\widehat{\bgamma}}^{[i]}=\{\bdelta\in\Delta^{[i]}: \bdelta'\widehat{\bQ}^{[i]}\bdelta-2(\widehat{\bgamma}^{[i]}-\bgamma^{[i]})'\bdelta\leq 0\}$ and $\Delta^{[i]}=\{\bdelta: \bdelta+\bw_0^{[i]}\in\mathcal{W}^{[i]}\}$.
Conditional on $\mathscr{H}$, $\widehat{\bgamma}^{[i]}$ is random, making the feasible set $\mathcal{M}_{\widehat{\bgamma}}^{[i]}$, and thus the resulting upper and lower bounds, stochastic as well. 
We can employ a normal distributional approximation of $\widehat{\bgamma}^{[i]}$ and, for instance, set the lower and upper bounds to 
$(\alpha_{\tin}/2)$-quantile of  
$\inf_{\bdelta\in\mathcal{M}_{\bG}^{[i]}}\bY_{\mathcal{N}(T_i+k)}'\bdelta$ and
$(1-\alpha_{\tin}/2)$-quantile of  
$\sup_{\bdelta\in\mathcal{M}_{\bG}^{[i]}}\bY_{\mathcal{N}(T_i+k)}'\bdelta$, respectively, conditional on $\mathscr{H}$,
where $\mathcal{M}_{\bG}^{[i]}=\{\bdelta\in\Delta^{[i]}:\bdelta'\widehat{\bQ}^{[i]}\bdelta-2\bG^{[i]\prime}\bdelta\leq 0\}$, 
$\bG^{[i]}|\mathscr{H}\thicksim\mathsf{N}(\mathbf{0}, \bSigma^{[i]})$ and  $\bSigma^{[i]}=\V[\widehat{\bgamma}^{[i]}|\mathscr{H}]$. It can be shown that, with high probability over $\mathscr{H}$,  such an interval has at least 
$(1-\alpha_{\tin}-\frac{c(\log T_0)^2}{\sqrt{T_0}})$ conditional coverage probability for the in-sample error
$\bY_{\mathcal{N}(T_i+k)}'(\widehat{\bw}^{[i]}-\bw_0^{[i]})$ for some constant $c>0$, where $\frac{c(\log T_0)^2}{\sqrt{T_0}}$ is some probability loss due to the normal approximation.

However, the above bounds cannot be directly implemented in practice, 
since the (centered) constraint set $\Delta^{[i]}$ depends on the unknown pseudo-true value $\bw_0^{[i]}$, and the normal vector $\bG^{[i]}$ depends on the unknown covariance matrix $\bSigma^{[i]}$. 
We thus propose a feasible simulation-based strategy allowing for \textit{possibly nonlinear} constraints, with unknown quantities replaced with plug-in approximations thereof.

First, we need a feasible constraint set $\Delta^{[i]\star}$ for simulation that is ``similar'' to $\Delta^{[i]}$ (near the origin). A general data-dependent strategy for constructing  $\Delta^{[i]\star}$ is discussed in Section \ref{subsec: define constraints in simulation}. We now focus on the L1-L2 constraint in our empirical application, where $\Delta^{[i]}$ consists of  one equality constraint ($\sum_{j\in\mathcal{N}}w_j=1$), a sequence of linear inequality constraints ($w_j\geq 0$ for $j\in\mathcal{N}$), and one \textit{nonlinear} inequality constraint ($\sum_{j\in\mathcal{N}}w_j^2\leq (Q^{[i]})^2$) on the weights vector $\bw=(w_j:j\in\mathcal{N})$. 
The equality constraint is maintained without adjustment in the simulation. For inequality constraints, we introduce a tuning parameter $\varrho^{[i]}$ to determine whether they are binding and then make adjustments accordingly.
Specifically, we set $\varrho^{[i]}=C^{[i]}\frac{\log T_0}{T_0}$ for some constant $C^{[i]}>0$, which can be viewed as a high-probability bound on the deviation of $\widehat{\bw}^{[i]}$ from $\bw_0^{[i]}$. A data-dependent way to choose $\varrho^{[i]}$ is described in Section \ref{subsec: define constraints in simulation}. Let $\widehat{\bw}=(\widehat{w}_j:j\in\mathcal{N})$. For each linear constraint $w_j\geq 0$, we consider it binding and use $w_j\geq \widehat{w}_j^{[i]}$ in the simulation if $\widehat{w}_j^{[i]}<\varrho^{[i]}$; otherwise, the original constraint, $w_j\geq 0$, is retained. For the nonlinear constraint $\|\bw\|_2^2\leq (Q^{[i]})^2$, impose $\|\bw\|_2^2\leq \|\widehat{\bw}^{[i]}\|_2^2+(\varrho^{[i]})^2$ in the simulation if $(Q^{[i]})^2-\|\widehat{\bw}^{[i]}\|_2^2<2\|\widehat{\bw}\|_2\varrho^{[i]}$, and retain $\|\bw\|_2^2\leq (Q^{[i]})^2$ otherwise.  Then,  $\Delta^{[i]\star}$ is defined as the set of vectors $\bw-\widehat{\bw}\in\mathbb{R}^{J_0}$ satisfying all these restrictions. 

It can be shown that, with high probability, all points in  $\Delta^{[i]}$  (near the origin) are contained in $\Delta^{[i]\star}$. In fact, if all constraints are linear in parameters, $\Delta^{[i]}$ and $\Delta^{[i]\star}$ are (locally) identical. With nonlinear constraints, the two sets are typically not identical but still remain close so that the ``enlargement'' from $\Delta^{[i]}$ to $\Delta^{[i]\star}$ is modest. See Section S.5.2 of the Supplemental Appendix for details. Given this fact,  searching for the supremum or infimum of $\bY_{\mathcal{N}(T_i+k)}'\bdelta$ over this (slightly) larger set $\Delta^{[i]\star}$ yields valid bounds on the in-sample error. 

Second, we need an estimator $\widehat{\bSigma}^{[i]}$ of the covariance matrix $\bSigma^{[i]}$. A variety of well-established heteroskedasticity/serial-correlation-robust estimators can be used. We require $\widehat{\bSigma}^{[i]}$ to be a ``good'' approximation of $\bSigma^{[i]}$ in the sense of condition (iii) in Corollary \ref{coro: basic setup} below. This allows us to approximate the infeasible normal distribution $\mathsf{N}(\bm{0},\bSigma^{[i]})$ by $\mathsf{N}(\bm{0},\widehat{\bSigma}^{[i]})$, which can be simulated using the data.

Once $\Delta^{[i]\star}$ and $\widehat{\bSigma}^{[i]}$ are available, we can simply draw random vectors from $\mathsf{N}(\bm{0}, \widehat{\bSigma}^{[i]})$ conditional on the data, and then set
\begin{equation}\label{eq: in-sample error bound, basic setup}
\myuline{M}_{\tin}=\underline{\mathfrak{c}}^\star(\alpha_{\tin}/2)
\quad\text{and}\quad
\myoline{M}_{\tin}=\bar{\mathfrak{c}}^\star(1-\alpha_{\tin}/2),
\end{equation}
where
\begin{align*}
\underline{\mathfrak{c}}^\star(\alpha_{\tin}/2)&=(\alpha_{\tin}/2)\text{-quantile of } 
\inf_{\bdelta\in\mathcal{M}_{\bG}^{[i]\star}}\bY_{\mathcal{N}(T_i+k)}'\bdelta,\\
\bar{\mathfrak{c}}^\star(1-\alpha_{\tin}/2)&=(1-\alpha_{\tin}/2)\text{-quantile of } \sup_{\bdelta\in\mathcal{M}_{\bG}^{[i]\star}}\bY_{\mathcal{N}(T_i+k)}'\bdelta,
\end{align*}
conditional on the data,
with 
$\mathcal{M}_{\bG}^{[i]\star}=\{\bdelta\in\Delta^{[i]\star}:\;
\bdelta'\widehat\bQ^{[i]}\bdelta-2(\bG^{[i]\star})'\bdelta\leq 0\}$ and 
$\bG^{[i]\star}|\mathsf{Data}\sim\mathsf{N}(\bm{0},\widehat\bSigma^{[i]})$.
Under some mild regularity conditions, 
$[\myuline{M}_{\tin},\, \myoline{M}_{\tin}]$ is a feasible prediction interval that achieves the desired conditional coverage of the counterfactual outcome $Y_{i(T_i+k)}$.

We are ready to construct bounds on the in-sample error for each treatment effect prediction.

\begin{examp}{TSUS}{Time-specific unit-specific predictand, continued}
    Since the in-sample error of $\widehat{\tau}_{ik}$ is the opposite of that of $\widehat{Y}_{i(T_i+k)}$, we simply set $\myuline{M}_{\tin}(\tau_{ik})=-\myoline{M}_{\tin}$ and $\myoline{M}_{\tin}(\tau_{ik})=-\myuline{M}_{\tin}$.
\end{examp}
   
\begin{examp}{TAUS}{Time-averaged unit-specific predictand, continued}
    Since the in-sample error of $\widehat{\tau}_{i\cdot}$ is (the opposite of) the average in-sample error of $\widehat{Y}_{i(T_i+k)}$ over time, we can set
    $\myuline{M}_{\tin}(\tau_{i\cdot})=\underline{\mathfrak{c}}^\star(\alpha_{\tin}/2)$ and
    $\myoline{M}_{\tin}(\tau_{i\cdot})=\bar{\mathfrak{c}}^\star(1-\alpha_{\tin}/2)$ where
    \begin{align*}
    \underline{\mathfrak{c}}^\star(\alpha_{\tin}/2)&=
    (\alpha_{\tin}/2)\text{-quantile of } 
    \inf_{\bdelta\in\mathcal{M}_{\bG}^{[i]\star}}
    -\frac{1}{T-T_i+1}\sum_{k=0}^{T-T_i}
    \bY_{\mathcal{N}(T_i+k)}'\bdelta,\\
    \bar{\mathfrak{c}}^\star(1-\alpha_{\tin}/2)&=(1-\alpha_{\tin}/2)\text{-quantile of } \sup_{\bdelta\in\mathcal{M}_{\bG}^{[i]\star}}
    -\frac{1}{T-T_i+1}\sum_{k=0}^{T-T_i}
    \bY_{\mathcal{N}(T_i+k)}'\bdelta,
    \end{align*}
    conditional on the data. 
\end{examp}
    
\begin{examp}{TSUA}{Time-specific unit-averaged predictand, continued}
    Since the in-sample error of $\widehat{\tau}_{\mathcal{Q}k}$ is (the opposite of) the average in-sample error of $\widehat{Y}_{i(T_i+k)}$ across multiple treated units, we can set
    $\myuline{M}_{\tin}(\tau_{\mathcal{Q}k})=\underline{\mathfrak{c}}^\star(\alpha_{\tin}/2)$ and
    $\myoline{M}_{\tin}(\tau_{\mathcal{Q}k})=\bar{\mathfrak{c}}^\star(1-\alpha_{\tin}/2)$ where
    \begin{align*}
    \underline{\mathfrak{c}}^\star(\alpha_{\tin}/2)&=
    (\alpha_{\tin}/2)\text{-quantile of } 
    \inf_{\bdelta\in\mathcal{M}_{\bG}^*}
    -\frac{1}{Q}\sum_{i:T_i\in\mathcal{Q}}\bY_{\mathcal{N}(T_i+k)}'\bdelta^{[i]},\\
    \bar{\mathfrak{c}}^\star(1-\alpha_{\tin}/2)&=(1-\alpha_{\tin}/2)\text{-quantile of } 
    \sup_{\bdelta\in\mathcal{M}_{\bG}^*}
    -\frac{1}{Q}\sum_{i:T_i\in\mathcal{Q}}\bY_{\mathcal{N}(T_i+k)}'\bdelta^{[i]},
    \end{align*}
    conditional on the data, and 
    $\mathcal{M}_{\bG}^\star=\{\bdelta\in\times_{i: T_i\in\mathcal{Q}}\;\Delta^{[i]\star}:
    \sum_{i:T_i\in\mathcal{Q}}[\bdelta^{[i]\prime}\widehat\bQ^{[i]}\bdelta^{[i]\prime}-2(\bG^{[i]\star})'\bdelta^{[i]\prime}]\leq 0\}$ with $\bdelta=(\bdelta^{[i]}: T_i\in\mathcal{Q})\in\mathbb{R}^{QJ_0}$. 
    In this case, the in-sample uncertainty depends on SC weights for multiple treated units, 
    so we aggregate the previously described quadratic constraint for each $\bdelta^{[i]}$ into one single quadratic constraint for the entire vector $\bdelta$.  
    With multiple treated units, we should draw the \textit{long} vector
    $\bG^\star:=(\bG^{[i]\star}: T_i\in\mathcal{Q})$ conditional on the data from
    $\mathsf{N}(\bm{0},\widehat\bSigma)$ with $\widehat\bSigma$ an estimate of $\bSigma=\V[\widehat{\bgamma}|\mathscr{H}]$ for $\widehat{\bgamma}=(\widehat{\bgamma}^{[i]}:T_i\in\mathcal{Q})$. 
    Therefore, the correlation structure among different treated units is implicitly captured.
\end{examp}
    
\begin{examp}{TAUA}{Time-averaged unit-averaged predictand, continued}
\hypertarget{exmp: taua, in-sample}{}
    Since the in-sample error of $\widehat{\tau}_{\cdot\cdot}$ is (the opposite of) the average in-sample error of $\widehat{Y}_{i(T_i+k)}$ across multiple treated units over time, we set 
    $\myuline{M}_{\tin}(\tau_{\cdot\cdot})=\underline{\mathfrak{c}}^\star(\alpha_{\tin}/2)$ and
    $\myoline{M}_{\tin}(\tau_{\cdot\cdot})=\bar{\mathfrak{c}}^\star(1-\alpha_{\tin}/2)$ where
    \begin{align*}
    \myuline{M}_{\tin}(\tau_{\cdot\cdot})&=
    (\alpha_{\tin}/2)\text{-quantile of } 
    \inf_{\bdelta\in\mathcal{M}_{\bG}^*}\;
    -\frac{1}{LJ_1}\sum_{k=1}^L\sum_{i\in\mathcal{E}}
    \bY_{\mathcal{N}(T_i+k)}'\bdelta^{[i]},\\
    \myoline{M}_{\tin}(\tau_{\cdot\cdot})&=(1-\alpha_{\tin}/2)\text{-quantile of } 
    \sup_{\bdelta\in\mathcal{M}_{\bG}^*}\;
    -\frac{1}{LJ_1}\sum_{k=1}^L\sum_{i\in\mathcal{E}}
    \bY_{\mathcal{N}(T_i+k)}'\bdelta^{[i]},
    \end{align*}
    conditional on the data, and $\mathcal{M}_{\bG}^\star=\{\bdelta\in\times_{i: i\in\mathcal{E}}\;\Delta^{[i]\star}:
    \sum_{i\in\mathcal{E}}[\bdelta^{[i]\prime}\widehat\bQ^{[i]}\bdelta^{[i]\prime}-2(\bG^{[i]\star})'\bdelta^{[i]\prime}]\leq 0\}$. 
    As in the previous example, we draw random vectors $\bG^\star:=(\bG^{[i]\star}: i\in\mathcal{E})$ conditional on the data from
    $\mathsf{N}(\bm{0},\widehat\bSigma)$ with $\widehat\bSigma$ an estimate of $\bSigma=\V[\widehat{\bgamma}|\mathscr{H}]$ for  
    $\widehat{\bgamma}=(\widehat{\bgamma}^{[i]}:i\in\mathcal{E})$.
\end{examp}

\begin{remark}[Alternative Bounds]\label{remark: alternative bounds}
    For the unit-averaged TSUA and TAUA predictands, we combine the quadratic constraints from optimization into a single constraint on the entire (centered) SC weights vector $\bdelta$. This ``aggregation'' strategy is consistent with the view of taking \eqref{eq: weights in basic framework} as a multiple-equation regression problem, as detailed in Section \ref{sec: theory}. 
    By contrast, we could also retain the individual quadratic constraints $\bdelta^{[i]\prime}\widehat\bQ^{[i]}\bdelta^{[i]\prime}-2(\bG^{[i]\star})'\bdelta^{[i]\prime}\leq 0$, for $T_i\in\mathcal{Q}$ or $i\in\mathcal{E}$, each restricting a subvector $\bdelta^{[i]}$ of $\bdelta$. This alternative strategy applies to the special case of SC analysis in which SC weights for each treated unit are constructed separately, yielding possibly tighter bounds on in-sample errors; see more detailed discussion in the Supplemental Appendix Section S.2.3.
    In Table S.2 of the Supplemental Appendix, we also illustrate the potential improvement of this method in our empirical application in terms of interval length.
\end{remark}

\begin{remark}[Scalable Optimization Implementations]\label{remark:Scalable Optimization Implementations}
    The proposed bounds on in-sample errors are suprema and infima of linear functions subject to some linear or quadratic constraints on parameters. In Section S.6 of the Supplemental Appendix, we show that such problems can be recast as conic optimization programs, which provide massive speed improvements in practice.
\end{remark}

\subsection{Out-of-Sample Uncertainty Quantification}\label{sec: out-of-sample error}

To bound the out-of-sample error, we propose an easy-to-implement approach based on non-asymptotic concentration inequalities. 
Recall that by Assumption \ref{assumption: dgp},
$u_{it}-\E[u_{it}|\mathscr{H}]$ is conditional-on-$\mathscr{H}$ sub-Gaussian with parameter $\sigma_{it}$, which implies that for any $\lambda>0$,
\begin{equation}\label{eq: subgaussian}
\P\Big(
|u_{it}-\E[u_{it}|\mathscr{H}]|\geq \lambda\Big|\mathscr{H}\Big)
    \leq 2\exp\Big(-\frac{\lambda^2}{2\sigma_{it}^2}\Big).
\end{equation}
Consequently, for the out-of-sample error $-u_{i(T_i+k)}$ of the counterfactual outcome prediction $\widehat{Y}_{i(T_i+k)}$, we can set
\begin{equation}\label{eq: out-of-sample error bound, basic setup}
\begin{split}
\myuline{M}_{\tout}&=-\E[u_{i(T_i+k)}|\mathscr{H}]-\sqrt{2\sigma_{i(T_i+k)}^2\log(2/\alpha_{\tout})}, 
	\quad\text{and}\\
\myoline{M}_{\tout}&=-\E[u_{i(T_i+k)}|\mathscr{H}]+\sqrt{2\sigma_{i(T_i+k)}^2\log(2/\alpha_{\tout})},
\end{split}
\end{equation}
which yields a prediction interval $[\myuline{M}_{\tout}, \, \myoline{M}_{\tout}]$ that covers $-u_{i(T_i+k)}$ with at least $(1-\alpha_{\tout})$ conditional coverage probability. 
This bound does not rely on the exact distribution of $u_{it}$ and is valid as long as $u_{it}$ has tails no fatter than those specified in \eqref{eq: subgaussian}. 
We emphasize that the sub-Gaussianity assumption is one of many possibilities. The above strategy could be applied using other concentration inequalities requiring weaker moment conditions, though the resulting prediction intervals may be wider.

In practice, one could first construct pre-treatment residuals $\widehat{u}_{it}=Y_{it}-\bY_{\mathcal{N}t}'\widehat{\bw}^{[i]}$ for each $i\in\mathcal{E}$ and $t<T_i$, and then 
estimate the conditional moments of $u_{it}$ employing various parametric or nonparametric regression of $\widehat{u}_{it}$. Such estimates can then be translated into the necessary estimates of $\E[u_{i(T_i+k)}|\mathscr{H}]$ and $\sigma_{i(T_i+k)}^2$ for constructing $\myuline{M}_{\tout}$ and $\myoline{M}_{\tout}$. The smallest parameter $\sigma_{it}^2$ that makes the tail probability bound \eqref{eq: subgaussian} hold is sometimes termed the optimal variance proxy, which is known to be the same as $\V[u_{it}|\mathscr{H}]$ if $u_{it}$ is \textit{strictly} sub-Gaussian conditional on $\mathscr{H}$ \citep{arbel2020strict,bobkov2024strictly}. In such cases estimating the first two conditional moments of $u_{it}$ suffices to implement the bounds in \eqref{eq: out-of-sample error bound, basic setup}. 
Moreover, the unknown conditional moments could also be set using external information, or tabulated across different values to assess the sensitivity of the resulting prediction intervals. 

We can apply the same idea to bound the out-of-sample errors of the four treatment effect predictions.

\begin{examp}{TSUS}{Time-specific unit-specific predictand, continued}
    Since the out-of-sample error in this case is $u_{it}$, we simply set 
    $\myuline{M}_{\tout}(\tau_{ik})=\E[u_{i(T_i+k)}|\mathscr{H}]-[2\sigma_{i(T_i+k)}^2\log(2/\alpha_{\tout})]^{1/2} $ and
    $\myoline{M}_{\tout}(\tau_{ik})=\E[u_{i(T_i+k)}|\mathscr{H}]+[2\sigma_{i(T_i+k)}^2\log(2/\alpha_{\tout})]^{1/2}$.
\end{examp}

\begin{examp}{TAUS}{Time-averaged unit-specific predictand, continued}
\hypertarget{exmp:taus, out-of-sample}{}
    Since $u_{it}-\E[u_{it}|\mathscr{H}]$ is condition-on-$\mathscr{H}$ sub-Gaussian with parameter $\sigma_{it}$, 
    it can be shown that the out-of-sample error $\mathtt{OutErr}(\tau_{i\cdot})$, as the average of $u_{it}$ over time, is also sub-Gaussian, satisfying that
    for any $\lambda>0$,
    \[
        \P\Big(|\mathtt{OutErr}(\tau_{i\cdot})-\E[\mathtt{OutErr}(\tau_{i\cdot})|\mathscr{H}]|\geq \lambda|\mathscr{H}\Big)\leq 2\exp\Big(-\frac{\lambda^2}{2\sigma_{i\cdot}^2}\Big), \quad 
        \sigma_{i\cdot}:=\frac{1}{T-T_i+1}\sum_{t=T_i}^{T}\sigma_{it}.
    \]
    Therefore, we can set
    \begin{align*}
    \myuline{M}_{\tout}(\tau_{i\cdot})&=\frac{1}{T-T_i+1}\sum_{k=0}^{T-T_i}\E[u_{i(T_i+k)}|\mathscr{H}]
    -\sqrt{2\sigma_{i\cdot}^2\log(2/\alpha_{\tout})},\quad \text{and}\\
    \myoline{M}_{\tout}(\tau_{i\cdot})&=\frac{1}{T-T_i+1}\sum_{k=0}^{T-T_i}\E[u_{i(T_i+k)}|\mathscr{H}]
    +\sqrt{2\sigma_{i\cdot}^2\log(2/\alpha_{\tout})}.
    \end{align*}
    This result holds regardless of the dependence structure of $u_{it}$, 
    but if $u_{it}$ is indeed independent over $t$, the same result holds with
    $\sigma_{i\cdot}:=\frac{1}{T-T_i+1}\big(\sum_{t=T_i}^T\sigma_{it}^2\big)^{1/2}$, leading to an improved bound.
\end{examp}

\begin{examp}{TSUA}{Time-specific unit-averaged predictand, continued}
    In this scenario, the out-of-sample error $\mathtt{OutErr}(\tau_{\mathcal{Q}k})$ is a cross-sectional average of $u_{it}$ at different times. 
    The uncertainty quantification strategy outlined previously in Example \hyperlink{exmp:taus, out-of-sample}{TAUS} can still be applied, with the caveat that it is uncommon in SC analysis to assume $u_{it}$ is stationary and/or independent over $i$. By contrast, it is reasonable to assume $u_{it}$ is stationary and/or independent (at least weakly dependent) over time. 
    Therefore, we employ the concentration inequality that holds under a general dependence structure and set
    \begin{align*}
    \myuline{M}_{\tout}(\tau_{\mathcal{Q}k})&=\frac{1}{Q}\sum_{i: T_i\in\mathcal{Q}}\E[u_{i(T_i+k)}|\mathscr{H}]
    -\sqrt{2\sigma_{\mathcal{Q}k}^2\log(2/\alpha_{\tout})},\quad \text{and}\\
    \myoline{M}_{\tout}(\tau_{\mathcal{Q}k})&=
    \frac{1}{Q}\sum_{i: T_i\in\mathcal{Q}}\E[u_{i(T_i+k)}|\mathscr{H}]
    +\sqrt{2\sigma_{\mathcal{Q}k}^2\log(2/\alpha_{\tout})}
    \end{align*}
    with $\sigma_{\mathcal{Q}k}:=\frac{1}{Q}\sum_{i:T_i\in\mathcal{Q}}\sigma_{i(T_i+k)}$.
\end{examp}

\begin{examp}{TAUA}{Time-averaged unit-averaged predictand, continued}
    Since the adoption time $T_i$ may be heterogeneous across $i$, the out-of-sample error $\mathtt{OutErr}(\tau_{\cdot\cdot})$ is an average of out-of-sample errors of different units in different periods, which is (conditionally) sub-Gaussian as well. Then, we set
    \begin{align*}
    \myuline{M}_{\tout}(\tau_{\cdot\cdot})&=\frac{1}{LJ_1}\sum_{k=1}^L\sum_{i\in\mathcal{E}}\E[u_{i(T_i+k)}|\mathscr{H}]
    -\sqrt{2\sigma_{\cdot\cdot}^2\log(2/\alpha_{\tout})},\quad \text{and}\\
    \myoline{M}_{\tout}(\tau_{\cdot\cdot})&=\frac{1}{LJ_1}\sum_{k=1}^L\sum_{i\in\mathcal{E}}\E[u_{i(T_i+k)}|\mathscr{H}]
    +\sqrt{2\sigma_{\cdot\cdot}^2\log(2/\alpha_{\tout})}
    \end{align*}
    with $\sigma_{\cdot\cdot}:=\frac{1}{LJ_1}
    \sum_{k=1}^K\sum_{i\in\mathcal{E}}\sigma_{i(T_i+k)}$.
\end{examp}

In addition to the concentration-based approach described above, other strategies, including location-scale models and quantile regression, were proposed in \citet*{Cattaneo-Feng-Titiunik_2021_JASA} for out-of-sample uncertainty quantification.  We briefly review them in the Supplemental Appendix, Section S.2.1.

\subsection{Algorithmic Implementation}

We summarize our methodology for the particular case illustrated in this section, focusing on the L1-L2 constraint and using $\tau_{ik}$ (TSUS) as the target predictand. A more detailed discussion of the procedures, including recommended rules of thumb for implementation and other practical regularization choices, can be found in Section S.7.2 of the Supplemental Appendix and in our companion software article \citet*[][Section 3.1 and Section 4.3]{Cattaneo-Feng-Palomba-Titiunik_2025_JSS}. In particular, the required optimization programs are implemented using modern, robust, and scalable conic programming methods (see Remark \ref{remark:Scalable Optimization Implementations} for details). Finally, all our proposed methods are readily available for implementation via our general-purpose software package \texttt{scpi} in \texttt{Python}, \texttt{R}, and \texttt{Stata}.

\begin{algorithm}[!ht]
    \caption{Uncertainty Quantification (Outcome-only Non-stationary Case)}\label{alg:unc quant}
    \begin{algorithmic}[1]
    \Require data $\{(Y_{it},T_i)_{i=1}^N:t=1,\ldots,T\}$, sets of donors $\mathcal{N}$ and treated units $\mathcal{E}$, confidence levels $\alpha_\tin,\alpha_\tout\in(0,1)$, and the number of simulations $S\in\mathbb{N}$.
    \Algphase{Step 1: Computation of $\widehat{\bw}$}
    \State tune  $Q^{[i]}$, for $ i\in\mathcal{E}$ (see Section S.7.2 of the Supplemental Appendix)
    \State solve the optimization problem in \eqref{eq: weights in basic framework} for each $i\in\mathcal{E}$
    \Algphase{Step 2: In-sample Uncertainty Quantification}
    \State form feasible constraint set $\Delta^{[i]\star}$ for each $i\in\mathcal{E}$
    \begin{algsubstates}
    \State tune $\varrho^{[i]}$ using \eqref{eq: tuning varrho} 
    \State define 
    \begin{align*}
        \Delta^{[i]\star}= \Big\{
        &\bw-\widehat{\bw}^{[i]}: \|\bw\|_1 =1, w_j\geq \widehat{w}_j^{[i]}\I(\widehat{w}_j^{[i]}<\varrho^{[i]}),\: j=1,\ldots, J_0,\\
        &\|\bw\|_2^2 - (Q^{[i]})^2\leq 
        (\|\widehat{\bw}^{[i]}\|_2^2 - (Q^{[i]})^2)\I(\|\bw^{[i]}\|_2^2 - ({Q}^{[i]})^2>-2\|\widehat{\bw}^{[i]}\|_2\varrho^{[i]}) +(\varrho^{[i]})^2 
        \Big\}
    \end{align*}
    \end{algsubstates}
    \State estimate $\bSigma^{[i]}$ for each $i\in\mathcal{E}$
    \begin{algsubstates}
    \State estimate pre-treatment residuals as $\widehat{u}_{it}=Y_{it}-\bY_{\mathcal{N}t}'\widehat{\bw}^{[i]}$ for $t<T_i$
    \State estimate $\E[u_{it}\mid\mathscr{H}]$ with a linear regression for $ t<T_i$
    \State estimate $\V[u_{it}\mid \mathscr{H}]$ with $\widehat{\V}[u_{it}\mid \mathscr{H}]=(\widehat{u}_{it}-\widehat{\E}[u_{it}\mid\mathscr{H}])^2$ for $ t<T_i$
    \State compute $\widehat{\bSigma}^{[i]}=\bY_\mathcal{N}^\prime \widehat{\mathbb{V}}[\mathbf{u}^{[i]}| \mathscr{H}]\bY_\mathcal{N})$ where $\bY_\mathcal{N}=[\bY_{\mathcal{N}1}\cdots \bY_{\mathcal{N}(T_i-1)}]^\prime$ 
    \For{$s \gets 1$ to $S$}:\setcounter{algsubstate}{0}
        \State draw $\mathbf{G}^{\star}\mid \mathscr{H} \sim \mathsf{N}(\mathbf{0}, \widehat{\boldsymbol{\Sigma}})$
        \State define $\ell^\star(\bdelta)= \bdelta^{\prime} \widehat{\mathbf{Q}} \bdelta-2\left(\mathbf{G}^{\star}\right)^{\prime} \bdelta$ with $\widehat{\mathbf{Q}}=\diag(\widehat{\mathbf{Q}}^{[i]}, i\in\mathcal{E}), \widehat{\bQ}^{[i]}=\bY_{\mathcal{N}}^\prime\bY_{\mathcal{N}}$
        \State solve
        \[l_{(s)} := \inf_{\substack{\bdelta \in\Delta^{[i]\star}\\ \ell^\star(\bdelta)\leq 0}} \bY_{\mathcal{N}(T_i+k)}'\bdelta\qquad\text{and}\qquad u_{(s)}:=\sup_{\substack{\bdelta \in\Delta^{[i]\star} \\ \ell^\star(\bdelta)\leq 0}}\bY_{\mathcal{N} (T_i+k)}'\bdelta\]
    \end{algsubstates}
    \State get $\underline{\mathfrak{c}}^\star$ as the $(\alpha_\tin/2)-$quantile of $\{l_{(s)}\}_{s=1}^S$ and $\bar{\mathfrak{c}}^\star$ as the $(1-\alpha_\tin/2)-$quantile of $\{u_{(s)}\}_{s=1}^S$
    \State get 
    $\myuline{M}_{\tin}=\underline{\mathfrak{c}}^\star, \myoline{M}_{\tin}=\bar{\mathfrak{c}}^\star$, and so $[\myuline{M}_{\tin}(\tau_{ik});\myoline{M}_{\tin}(\tau_{ik})]= [-\myoline{M}_{\tin};-\myuline{M}_{\tin}]$
    \Algphase{Step 3: Out-of-sample Uncertainty Quantification}
    \State use pre-treatment residuals to estimate $\mathbb{E}[u_{i(T_i+k)}| \mathscr{H}]$ and $\sigma_{i(T_i+k)}^2$
    \State get
    $[\myuline{M}_{\tout}(\tau_{ik});\myoline{M}_{\tout}(\tau_{ik})]=\widehat{\mathbb{E}}\left[u_{i(T_i+k)}| \mathscr{H}\right]\pm\sqrt{2 \widehat\sigma_{i(T_i+k)}^2 \log \left(2 / \alpha_\tout\right)} $ 
    \Algphase{Step 4: Compute Prediction Intervals}
    \State form an interval as $[\widehat{\tau}_{ik}-\myoline{M}_{\tin}(\tau_{ik})-\myoline{M}_{\tout}(\tau_{ik});\widehat{\tau}_{ik}-\myuline{M}_{\tin}(\tau_{ik})-\myuline{M}_{\tout}(\tau_{ik})]$
    \end{algorithmic}
\end{algorithm}

\subsection{Theoretical Justification}\label{sec: theoretical justification}

The following corollary,  as a special case of our more general results presented in Section \ref{sec: theory}, provides a theoretical justification of the proposed method, which closely matches our empirical application. 
Let $T_0=T_1-1$, and $m_\ell^{[i]}(\bw)\leq 0$ be the $\ell$-th inequality constraint among $-w_j\leq 0$, $j\in\mathcal{N}$, and $\|\bw\|_2^2\leq (Q^{[i]})^2$. Define $\tilde{\bQ}=\frac{1}{T_0}\sum_{t=1}^{T_0}\tilde\bG(t/T_0)\tilde\bG(t/T_0)'$ where $\tilde\bG$ is a mean-zero Brownian motion on $[0,1]$ with variance $\E[\bv_{t}\bv_{t}']$. We use $s_{\min}(\bA)$ and $s_{\max}(\bA)$ to denote the minimum and maximum singular values of a matrix $\bA$, respectively.  
Also, for simplicity, in the following corollary we define $\bSigma=\V[\widehat{\bgamma}|\mathscr{H}]$ for $\widehat{\bgamma}=(\widehat{\bgamma}^{[i]}: i\in\mathcal{E})$ and let $\widehat{\bSigma}$ be an estimator of $\bSigma$, as in Example \hyperlink{exmp: taua, in-sample}{TAUA}.

\begin{coro}\label{coro: basic setup}
Let Assumption \ref{assumption: dgp} hold. In addition, suppose that
with probability over $\mathscr{H}$ at least $1-\pi_0$, the following conditions hold:
 \begin{enumerate}[label=\normalfont(\roman*),noitemsep]
     \item $\min_{1\leq t\leq T}s_{\min}(\E[\bu_{t}\bu_t'|\mathscr{H}])>\eta$ for some constant $\eta>0$;
     \item $(\log T_0)^{-1/5}\leq s_{\min}(\tilde\bQ)\leq 
     s_{\max}(\tilde\bQ)\leq 
     (\log T_0)^{1/5}$;
     \item $\P(\|\widehat{\bSigma}-\bSigma\|\leq\epsilon_{\Sigma,1}^\star)|\mathscr{H})\geq 
     1-\epsilon^\star_{\Sigma,2}$ with 
     $\epsilon_{\Sigma,1}^\star\leq \frac{T_0^2\eta}{8\sqrt{d}(\log T_0)^{1/5}}$;
     \item For all $1\leq \ell\leq J_0+1$, 
    $\varrho_\ell^{[i]}<|m_{\ell}^{[i]}(\bw_0^{[i]})|-(\log T_0)T_0^{-1}$ if
    $m_{\ell}^{[i]}(\bw_0^{[i]})\neq 0$.
 \end{enumerate}

    Then, for any causal prediction $\widehat\tau\in\{\widehat{\tau}_{ik}, \widehat{\tau}_{i\cdot}, \widehat{\tau}_{\mathcal{Q}k}, \widehat{\tau}_{\cdot\cdot}\}$, when $T_0$ is large enough, 
	\[
	\P\Big\{\P\big(
        \tau\in[
	\widehat{\tau}-\myoline{M}_{\tin}(\tau)-\myoline{M}_{\tout}(\tau),\;
	\widehat{\tau}-\myuline{M}_{\tin}(\tau)-\myuline{M}_{\tout}(\tau)]
    \big|\mathscr{H}\big)\geq 1-\alpha_{\tin}-\alpha_{\tout}-\epsilon\Big\}\geq 1-\pi,
	\]
 where $\epsilon=2\mathfrak{C}_\epsilon(\log T_0)^2T_0^{-1/2}+
 4\epsilon_{\Sigma,1}^\star\sqrt{d}
 (\log T_0)^{1/5}T_0^{-2}\eta^{-1}+\epsilon_{\Sigma,2}^\star+3T_0^{-1}$,
 $\pi=\pi_0+\frac{\mathfrak{C}_\pi}{T_0}$, and 
 $\mathfrak{C}_\epsilon$ and $\mathfrak{C}_\pi$ are some constants characterized in the proof.
\end{coro}

The additional technical conditions imposed in this corollary are intuitive: (i) is a mild restriction used to guarantee the variance matrix $\bSigma$ is non-degenerate;  
(ii) is used to characterize the upper and lower bounds on the matrices $\widehat{\bQ}^{[i]}$ and can be shown to hold with high probability when $T_0$ is large; 
(iii) is a high-level condition on the convergence of the variance estimator $\widehat{\bSigma}$ and can be verified on a case-by-case basis; and (iv) guarantees that the non-binding constraints can be differentiated from the binding ones under our thresholding rule, making the inclusion restriction on the constraint sets imposed in Theorem \ref{thm: coverage error approximation, plug-in} satisfied.
If these conditions hold with high probability ($\pi_0$ is small), then
this corollary shows that the proposed prediction intervals can achieve approximately $(1-\alpha_{\tin}-\alpha_{\tout})$ conditional coverage probability, with the probability loss $\epsilon$ decreasing as $T_0$ grows and $\epsilon_{\Sigma,1}^\star$ and $\epsilon_{\Sigma,2}^\star$ get small (i.e., the variance estimator $\widehat{\bSigma}$ becomes more ``precise'').

\subsection{Simultaneous Prediction Intervals}\label{sec: simultaneous PI}

So far we have focused on constructing prediction intervals that have high coverage of the desired predictands or treatment effects, in particular, the TSUS predictand, which captures the effect of the intervention for a given unit in \textit{each} post-treatment period. In some applications, it might be appealing to construct prediction intervals that have high \textit{simultaneous} coverage in multiple post-treatment periods, usually termed \textit{simultaneous prediction intervals}. They can be employed to test, for example, whether the largest (or smallest) treatment effect across different periods is significantly different from zero.

Specifically, for a particular treated unit $i\in\mathcal{E}$, we aim to construct a sequence of intervals $\mathcal{I}_k$ for $0\leq k\leq L$ for some $L\leq T-T_i$ such that
\[
  \P\Big\{\P\big[\tau_{ik}\in\mathcal{I}_k, \text{ for all } 0\leq k\leq L\,\big|\, \mathscr{H}\big]\geq 1-\alpha\Big\}\geq 1-\pi.
\]
As described before, the uncertainty of the predicted TSUS effect $\widehat{\tau}_{ik}$ comes from the in-sample error 
$\mathsf{InErr}(\tau_{ik})=-Y_{\mathcal{N}(T_i+k)}'(\widehat{\bw}^{[i]}-\bw_0^{[i]})$ and 
the out-of-sample error $\mathsf{OutErr}(\tau_{ik})=u_{i(T_i+k)}$. 

Regarding the in-sample error, the following is an immediate generalization of the bound described in Section \ref{sec: in-sample error}, which enjoys simultaneous coverage in multiple periods: 
\begin{equation}
\myuline{M}_{\tin}(\tau_{ik})=\underline{\mathfrak{c}}^\star(\alpha_{\tin}/2)
\qquad\text{and}\qquad
\myoline{M}_{\tin}(\tau_{ik})=\bar{\mathfrak{c}}^\star(1-\alpha_{\tin}/2),
\end{equation}
where
\begin{align*}
\underline{\mathfrak{c}}^\star(\alpha_{\tin}/2)&= (\alpha_{\tin}/2)\text{-quantile  of }
\inf_{\bdelta\in\mathcal{M}_{\bG}^{[i]\star}, \, 0\leq k\leq L}\;
-\bY_{\mathcal{N}(T_i+k)}'\bdelta,\\
\bar{\mathfrak{c}}^\star(1-\alpha_{\tin}/2)&=(1-\alpha_{\tin}/2)\text{-quantile  of } 
\sup_{\bdelta\in\mathcal{M}_{\bG}^{[i]\star}, \, 0\leq k\leq L}\;
-\bY_{\mathcal{N}(T_i+k)}'\bdelta,
\end{align*}
conditional on the data. It can shown that this generally guarantees that 
$[\myuline{M}_{\tin}(\tau_{ik}),\; \myoline{M}_{\tin}(\tau_{ik})]$ can simultaneously cover the in-sample error $\mathsf{InErr}(\tau_{ik})$ for all $0\leq k\leq L$, with $\P(\cdot|\mathscr{H})$-probability at least $1-\alpha_{\tin}$, up to some small loss.

Regarding the out-of-sample error, an easy-to-implement strategy analogous to that described in Section \ref{sec: out-of-sample error} is to adjust the bounds on the out-of-sample error $\mathsf{OutErr}(\tau_{ik})=u_{i(T_i+k})$ based on maximal inequalities. 
Recall that each $u_{i(T_i+k)}-\E[u_{i(T_i+k)}|\mathscr{H}]$, $0\leq k\leq L$, is conditionally sub-Gaussian with parameter $\sigma_{i(T_i+k)}$ (but is not necessarily independent over $k$). Then, for any $\lambda> 0$,
$$
\P\Big(\max_{0\leq k\leq L}|u_{i(T_i+k)}-\E[u_{i(T_i+k)}|\mathscr{H}]|\geq \lambda\Big|\mathscr{H}\Big)
\leq 2\sum_{k=0}^{L}\exp\Big(-\frac{\lambda^2}{2\sigma_{i(T_i+k)}^2}\Big).
$$
If $\sigma_{i(T_i+k)}\leq \sigma_i$ for all $0\leq k\leq T-T_i$, then one can set
$\myuline{M}_{\tout}(\tau_{ik})=
\E[u_{i(T_i+k)}|\mathscr{H}]-\lambda$ and 
$\myoline{M}_{\tout}(\tau_{ik})=\E[u_{i(T_i+k)}|\mathscr{H}]+\lambda$ with $\lambda=\sqrt{2\sigma_{i}^2\log (2(L+1)/\alpha_2)}$.
It can be shown that $[\myuline{M}_{\tout}(\tau_{ik}),\; \myoline{M}_{\tout}(\tau_{ik})]$ can simultaneously cover $\mathsf{OutErr}(\tau_{ik})$ for all $0\leq k\leq L$, with $\P(\cdot|\mathscr{H})$-probability at least $1-\alpha_{\tout}$.
Compared with prediction intervals with validity for each period constructed similarly, these simultaneous prediction intervals are slightly wider due to the additional factor $\sqrt{\log (L+1)}$. 
In practice, assuming $u_{it}$'s are strictly sub-Gaussian, one only needs to estimate the conditional mean and variance of $u_{it}$ using the pre-treatment residuals to implement the proposed bounds; flexible parametric or non-parametric estimation methods can be used.

Again, the sub-Gaussianity assumption can be relaxed using other concentration inequalities requiring weaker moment conditions, though the resulting simultaneous prediction intervals may be wider. 
Also, there are other strategies to construct prediction intervals that simultaneously cover multiple out-of-sample errors, though they are computationally more cumbersome and usually require more stringent conditions. See Section S.2.2 of the Supplemental Appendix for a brief discussion. 

The idea outlined above to achieve simultaneous coverage is general and can also be used to, for example, construct prediction intervals that simultaneously cover the TSUS predictands for multiple treated units rather than for multiple post-treatment periods. In our empirical application, we construct simultaneous prediction intervals for time-averaged effects across different economies; see details in Section \ref{sec: empirical results}.

\section{Empirical Results}\label{sec: empirical results}

We use our framework to evaluate the effect of economic liberalization on (log) real GDP per capita in the sixteen countries that adopted economic liberalization in Sub-Saharan Africa: Benin, Botswana, Cabo Verde, Cameroon, Gambia, Ghana, Guinea, Guinea, Guinea-Bissau, Ivory Coast, Mali, Mauritius, Niger, South Africa, Uganda, and Zambia. 

We emphasize that when the target predictand does not involve an average across countries (e.g., TSUS and TAUS), we include not-yet-treated countries in the donor pool. Including not-yet-treated units in other scenarios would raise issues, as their pre-treatment features would appear both in the conditioning set $\mathscr{H}$ and in the outcome vector, thereby invalidating the attempt of quantifying their contribution to prediction uncertainty.

Overall, our point predictions suggest that (external) liberalization episodes in Sub-Saharan Africa had no impact on GDP per capita. Although we find that the point prediction for Botswana is large and positive, our predictands and uncertainty characterization indicate that this predicted effect cannot be distinguished from zero. Our new evidence thus suggests that Botswana's success story was possibly the cause rather than the consequence of economic liberalization. In what follows, we report the predictands described in Section \ref{sec: basic framework}, Examples \hyperlink{exmp:tsus}{TSUS}, \hyperlink{exmp:taus}{TAUS}, and \hyperlink{exmp:tsua}{TSUA}.

In all our results, we compute the weights using an L1-L2 constraint and use pre-treatment GDP as the only feature. In the Supplemental Appendix Section S.8, we also report the results using (\textit{i}) a simplex-type constraint, (\textit{ii}) a ridge-type constraint, (\textit{iii}) more than one feature, (\textit{iv}) using placebo treatment dates, and (\textit{v}) leaving one of the donors out at a time. All the implementation details are reported in Supplemental Appendix Section S.7.

\medskip
\noindent\uline{TSUS predicted effects in every period after liberalization ($\tau_{ik}$)}. We first analyze the predicted treatment effect for every individual country that adopts liberalization in each post-liberalization period (up to five years after adoption), which is an example of the \hyperlink{exmp:tsus}{TSUS} predictand. In Figure \ref{fig: individual treatment effects}, we show the predicted synthetic control outcomes (panel (a)) with the corresponding 90\% prediction intervals, and the estimated weights $\widehat{\bw}^{[i]}, i\in\mathcal{E}$ (panel (b)) for our sixteen countries. 

\begin{figure}[!ht]
    \centering
    \caption{Time-specific unit-specific (TSUS) predicted effects in every period, $\widehat{\tau}_{ik}$.}
    \label{fig: individual treatment effects}
     \begin{subfigure}[b]{0.98\textwidth}
         \centering
         \resizebox{0.95\textwidth}{!}{%
         \input{figs/Africa_WaveAll_indiv_L1-L2.tex}
         }
         \caption{$Y_{it}(T_i)$ and $\widehat{Y}_{it}(\infty)$}
         \label{subfig: individual treatment effects series}
     \end{subfigure}
    \begin{subfigure}[b]{0.98\textwidth}
         \centering
         \resizebox{0.95\textwidth}{!}{%
         \input{figs/Africa_WaveAll_indiv_L1-L2_w.tex}
         }
         \caption{$\widehat{w}_{j}^{[i]}, i\in\mathcal{E},j\in\mathcal{N}$}
     \end{subfigure} 
 \par
	\begin{center}
		\parbox[1]{\textwidth}{\footnotesize \textit{Notes:} Panel (a): TSUS prediction for every country in each of five periods after treatment. Blue bars report 90\% prediction intervals. In-sample uncertainty is quantified using 200 simulations, whereas out-of-sample uncertainty is quantified using sub-Gaussian bounds. Panel (b): each dot represents the weight that the donor (row) gets in forming the synthetic control for the treated unit (column). When there is no dot, it means that the unit was not part of the donor pool for the treated unit in question.}
	\end{center}     
\end{figure}

In most countries, the realized trajectory of GDP per capita (black lines) lies roughly on top of the synthetic one (blue lines), suggesting that in the absence of the liberalization event, real income per capita would have been approximately equal. The 90\% prediction intervals (blue vertical bars) indicate that, in almost all cases, the distance between the actual GDP series and the counterfactual one is not different from zero with high probability for almost all units and periods. 

For Botswana, Guinea, and Mauritius the realized trajectories of GDP per capita (black lines) lay above the synthetic ones, suggesting a positive treatment effect. The 90\% prediction intervals show that this effect can be distinguished from zero with high probability after three years from the liberalization (with the exception of Botswana after four years), whereas it cannot in the first two years. We stress that the pre-treatment fit for Botswana is poorer than the one for Mauritius and Guinea. In Section S.8 of the Supplemental Appendix, we show that these findings are robust when a simplex-type constraint is employed, but not with a ridge-type constraint or when a second feature (the investment-to-GDP ratio) is included.

\medskip
\noindent\uline{TAUS predicted effects, averaged over five years ($\tau_{i\cdot}$)}. The second causal predictand of interest is the effect for each of the sixteen African countries we study, averaged over the five periods following the liberalization treatment in each country (up to the year 2000). This is an example of the \hyperlink{exmp:TAUS}{TAUS} predictand. Figure \ref{fig: unit treatment effects} shows that in almost all countries the liberalization episode seems to have a negligible effect on real GDP per capita.

By looking at the prediction intervals, we can see that 13 treated units show an average effect that cannot be distinguished from zero with high probability. On the other hand, for Botswana, Guinea, and Mauritius we find a positive TAUS which is also statistically significant with high probability. Again, we stress the poor pre-treatment fit in interpreting the results for Botswana. Moreover, in Section S.8 of the Supplemental Appendix, we show that employing a simplex-type constraint yields similar results, although these results are not robust when a second feature (investment-to-GDP ratio) is introduced or when a ridge-type constraint is used.

\begin{figure}[!ht]
    \centering
    \caption{Time-averaged unit-specific (TAUS) predicted effects, averaged over five years, $\widehat{\tau}_{i\cdot}$.}
    \label{fig: unit treatment effects}
     \begin{subfigure}[b]{0.98\textwidth}
         \centering
         \resizebox{\textwidth}{!}{%
         \input{figs/Africa_WaveAll_unit_L1-L2.tex}}
         \caption{$Y_{it}(T_i)$ and $\widehat{Y}_{it}(\infty)$}
         \label{subfig: unit treatment effects series}
     \end{subfigure}\\
    \begin{subfigure}[b]{0.98\textwidth}
         \centering
         \resizebox{\textwidth}{!}{%
         \input{figs/Africa_WaveAll_unit_L1-L2_w.tex}}
         \caption{$\widehat{w}_{j}^{[i]}, i\in\mathcal{E},j\in\mathcal{N}$}
     \end{subfigure} 
 \par
	\begin{center}
		\parbox[1]{\textwidth}{\footnotesize \textit{Notes:} Panel (a): TAUS prediction for every country averaged over the five periods following treatment (up to the year 2000). Blue bars report 90\% prediction intervals. In-sample uncertainty is quantified using 200 simulations, whereas out-of-sample uncertainty is quantified using sub-Gaussian bounds. Panel (b): each dot represents the weight that the donor (row) gets in forming the synthetic control for the treated unit (column). When there is no dot, it means that the unit was not part of the donor pool for the treated unit in question.}
	\end{center}     
\end{figure}

\medskip
\noindent\uline{TSUA predicted effects, averaged over countries that liberalized in each of three waves: before 1987, between 1987 and 1991, and after 1991 ($\tau_{\mathcal{Q}_1k}$, $\tau_{\mathcal{Q}_2k}$, $\tau_{\mathcal{Q}_3k}$)}.
In interpreting the evidence for each individual country, \cite{Billmeier-Nannicini_2013_RESTAT} consider the hypothesis that liberalization only led to economic growth for countries that liberalized early. Our framework allows us to group countries according to the era in which they liberalized and consider their joint trajectory in a formal way. To do so, we use the three waves considered by \cite{Billmeier-Nannicini_2013_RESTAT} and study three different predictands that average over all countries that liberalized during each wave: the TSUA effect for all countries that liberalized before 1986 (Botswana, Gambia, Ghana, and Guinea), $\tau_{\mathcal{Q}_1k}$ with $\mathcal{Q}_1=\{t: t<1987\}$; the TSUA effect for all countries that liberalized between 1987 and 1991 (Benin, Cabo Verde, Guinea-Bissau, Mali, South Africa, and Uganda), $\tau_{\mathcal{Q}_2k}$ with $\mathcal{Q}_2=\{t: 1987\leq t\leq 1991\}$; and the TSUA  effect for all countries that liberalized between 1992 and 1994 (Burkina Faso, Burundi, Cameroon, Ethiopia, Ivory Coast, Mozambique, Niger, Tanzania, and Zambia), $\tau_{\mathcal{Q}_3k}$ with $\mathcal{Q}_3=\{t: 1991< t\leq 1994\}$. The predicted effect is calculated for every post-liberalization period (up to five years), each representing a specific example of the \hyperlink{exmp:tsua}{TSUA} predictand.

The results, presented in Figure \ref{fig: time treatment effects waves}, show that, with high probability and when joint prediction intervals are considered, we can conclude that countries that liberalized before 1987 had a positive TSUA predicted effect. We caution the reader in interpreting this evidence because of the poor pre-treatment fit in Figure \ref{fig: time treatment effects waves}(a). The quality of the pre-treatment fit follows from the small cardinality of the donor group for this predictand, given the exclusion of not-yet-treated countries. These findings are robust to the inclusion of a second feature and if a simplex-type constraint is employed (see Section S.8 of the Supplemental Appendix). However, relying on a ridge-type constraint (Section S.8.1 of the Supplemental Appendix), which greatly improves the pre-treatment fit and hence is our preferred specification for this predictand, shows that, with high probability, we can conclude that countries that liberalized enjoyed a positive TSUA effect after 5 years in all three waves, thus not supporting the original hypothesis of \cite{Billmeier-Nannicini_2013_RESTAT}.

\begin{figure}[!ht]
    \centering
    \medskip
    \caption{Time-specific unit-averaged (TSUA) predicted effects in each period, averaged over three groups of countries.}
        \label{fig: time treatment effects waves}
\centering \texttt{\uline{Countries that Liberalized Before 1987}, $\widehat{\tau}_{\mathcal{Q}_1 k}$}
\begin{minipage}[t]{1\textwidth}
     \begin{subfigure}[b]{0.48\textwidth}
         \centering
          \resizebox{0.95\textwidth}{!}{%
         \input{figs/Africa_Wave1_time_L1-L2.tex}
         }
         \caption{$Y_{it}(T_i)$ and $\widehat{Y}_{it}(\infty)$}
         \label{subfig: time treatment effects series before 1987}
     \end{subfigure}
    \begin{subfigure}[b]{0.48\textwidth}
         \centering
         \resizebox{0.95\textwidth}{!}{%
         \input{figs/Africa_Wave1_time_L1-L2_w.tex}
         }
         \caption{$\widehat{w}_{j}^{[i]}, i\in\mathcal{E},j\in\mathcal{N}$}
     \end{subfigure}\\
    \end{minipage}
 \centering \texttt{\uline{Countries that Liberalized in 1987-1991}, $\widehat{\tau}_{\mathcal{Q}_2 k}$}
 \begin{minipage}[t]{1\textwidth}     
      \begin{subfigure}[b]{0.48\textwidth}
         \centering
         \resizebox{0.95\textwidth}{!}{%
         \input{figs/Africa_Wave2_time_L1-L2.tex}
         }
         \caption{$Y_{it}(T_i)$ and $\widehat{Y}_{it}(\infty)$}
         \label{subfig: time treatment effects series in 1987-1991}
     \end{subfigure}
     \hfill
    \begin{subfigure}[b]{0.48\textwidth}
         \centering
         \resizebox{0.95\textwidth}{!}{%
         \input{figs/Africa_Wave2_time_L1-L2_w.tex}
         }
         \caption{$\widehat{w}_{j}^{[i]}, i\in\mathcal{E},j\in\mathcal{N}$}
     \end{subfigure}\\
     \end{minipage}
\centering  \texttt{\uline{Countries that Liberalized after 1991}, $\widehat{\tau}_{\mathcal{Q}_3 k}$}
 \begin{minipage}[t]{1\textwidth} 
      \begin{subfigure}[b]{0.48\textwidth}
         \centering
         \resizebox{0.95\textwidth}{!}{%
         \input{figs/Africa_Wave3_time_L1-L2.tex}
         }
         \caption{$Y_{it}(T_i)$ and $\widehat{Y}_{it}(\infty)$}
         \label{subfig: time treatment effects series after 1991}
     \end{subfigure}
     \hfill
    \begin{subfigure}[b]{0.48\textwidth}
         \centering
         \resizebox{0.95\textwidth}{!}{%
         \input{figs/Africa_Wave3_time_L1-L2_w.tex}
         }
         \caption{$\widehat{w}_{j}^{[i]}, i\in\mathcal{E},j\in\mathcal{N}$}
     \end{subfigure}
     \end{minipage}
 \par
	\begin{center}
		\parbox[1]{\textwidth}{\footnotesize \textit{Notes:} TSUA prediction in every period after treatment (up to five years), averaged over all countries that liberalized in each of three waves: before 1987 (Botswana, Gambia, Ghana, and Guinea), between 1987 and 1991 (Benin, Cabo Verde, Guinea-Bissau, Mali, South Africa, and Uganda), and after 1991 (Burkina Faso, Burundi, Cameroon, Ethiopia, Ivory Coast, Mozambique, Niger, Tanzania, and Zambia). Blue bars report 90\% prediction intervals, whereas blue-shaded areas report 90\% simultaneous prediction intervals. In-sample uncertainty is quantified using 200 simulations, whereas out-of-sample uncertainty is quantified using sub-Gaussian bounds. Panel (b): each dot represents the weight that the donor (row) gets in forming the synthetic control for the treated unit (column). When there is no dot, it means that the unit was not part of the donor pool for the treated unit in question.}
	\end{center}     
\end{figure}

\medskip
\noindent\uline{TSUA predicted effects, averaged over all liberalized countries ($\tau_{\mathcal{E} k}$)}. In this last and fourth exercise, we focus on a popular causal predictand: the effect in every period after treatment averaged over all the treated countries. This is yet another example of the \hyperlink{exmp:tsua}{TSUA} predictand, with $\mathcal{Q}=\mathcal{E}$. This predictand thus averages over all countries that liberalized, regardless of when they did so. We report this predictand in every year after the adoption of liberalization (up to five years) which occurs at different times for different countries. Figure \ref{fig: time treatment effects} reports the results. To compute this predictand, we first construct a synthetic control for each treated unit (see panel (b)) and then we pool all the synthetic controls together in every post-treatment period to get a single prediction and a single prediction interval. Panel (a) and panel (b) show that pooling across the 16 African countries that embarked on liberalization programs helps reduce the uncertainty surrounding the synthetic trajectory. This TSUA predictand shows that, with high probability, the liberalization had a statistically significant impact on GDP-per-capita after two years. This result is robust to both simplex- and ridge-type constraints (this latter also with a better pre-treatment fit), but not to the inclusion of a second feature (see Section S.8 of the Supplemental Appendix).

\begin{figure}[!ht]
    \centering
    \caption{Time-specific unit-averaged (TSUA) predicted effect, averaged over all treated units, $\widehat{\tau}_{\mathcal{E}k}$.}
    \label{fig: time treatment effects}
     \begin{subfigure}[b]{0.98\textwidth}
         \centering
         \resizebox{0.95\textwidth}{!}{%
         \input{figs/Africa_WaveAll_time_L1-L2.tex}
         }             
         \caption{$Y_{it}(T_i)$ and $\widehat{Y}_{it}(\infty)$}
         \label{subfig: time treatment effects series}
     \end{subfigure} \\
    \begin{subfigure}[b]{0.98\textwidth}
         \centering
         \resizebox{0.95\textwidth}{!}{%
         \input{figs/Africa_WaveAll_time_L1-L2_w.tex}
         }         
         \caption{$\widehat{w}_{j}^{[i]}, i\in\mathcal{E},j\in\mathcal{N}$}
     \end{subfigure} 
 \par
	\begin{center}
		\parbox[1]{\textwidth}{\footnotesize \textit{Notes:} Panel (a): TSUA prediction in every period after treatment, averaged over all the treated countries. Blue bars report 90\% prediction intervals, whereas blue-shaded areas report 90\% simultaneous prediction intervals. In-sample uncertainty is quantified using 200 simulations, whereas out-of-sample uncertainty is quantified using sub-Gaussian bounds. Panel (b): each dot represents the weight that the donor (row) gets in forming the synthetic control for the treated unit (column). When there is no dot, it means that the unit was not part of the donor pool for the treated unit in question.}
	\end{center}     
\end{figure}

\section{Theoretical Foundations}\label{sec: theory}

This section presents a general framework that can accommodate more flexible specifications in SC analysis. As in Section \ref{sec: basic framework}, we still consider the case with $J_0$ never-treated units and $J_1$ ever-treated units that adopt a treatment at possibly different times.
However, now we assume that a user may want to obtain SC weights by matching on $M$ features (denoted by a subscript $l=1, \cdots, M$ below) with additional covariates adjustment, rather than relying solely on pre-treatment outcomes.

Specifically, let $\bA_l^{[i]}=(a_{1,l}^{[i]}, \cdots, a_{T_{i0},l}^{[i]})'\in\mathbb{R}^{T_{i0}}$ be the $l$-th feature of the treated unit $i$ measured in $T_{i0}$ (user-specified) pre-treatment periods. For each feature $l$ and each treated unit $i$, there exist $J_0+K$ variables that are used to predict or match the $T_{i0}$-dimensional vector $\bA_l^{[i]}$. These $J_0+K$ variables are separated into two groups denoted by $\bB_l^{[i]}=(\bB_{1,l}^{[i]}, \bB_{2,l}^{[i]}, \cdots, \bB_{J_0, l}^{[i]})\in\mathbb{R}^{T_{i0}\times J_0}$ and $\bC_l^{[i]}=(\bC_{1,l}^{[i]}, \cdots, \bC_{K,l}^{[i]})\in\mathbb{R}^{T_{i0}\times K}$, respectively. 
More precisely, for each $j=1, \ldots, J_0$, $\bB_{j,l}^{[i]}=(b_{j1,l}^{[i]}, \cdots, b_{jT_{i0},l}^{[i]})'$ corresponds to the $l$-th feature of the $j$-th unit in the donor pool measured in $T_{i0}$ pre-treatment periods, and for each $k=1, \ldots, K$, $\bC_{k,l}^{[i]}=(c_{k1,l}^{[i]}, \cdots, c_{kT_{i0},l}^{[i]})'$ is another vector of control variables used to predict $\bA_l^{[i]}$ over the same pre-intervention time span. 
Stacking the $M$ equations (corresponding to $M$ features) for each treated unit, we define
\[\underbrace{\mathbf{A}^{[i]}}_{T_{i0}\cdot M\times 1} = \begin{bmatrix} \mathbf{A}_1^{[i]} \\ \vdots \\ \mathbf{A}_M^{[i]} \end{bmatrix}, \quad \underbrace{\mathbf{B}^{[i]}}_{T_{i0}\cdot M \times J_0} = \begin{bmatrix} \mathbf{B}_1^{[i]} \\ \vdots \\ \mathbf{B}_M^{[i]} \end{bmatrix},\quad \underbrace{\mathbf{C}^{[i]}}_{T_{i0}\cdot M \times K\cdot M}=\begin{bmatrix} \mathbf{C}_1^{[i]} & \mathbf{0} & \cdots & \mathbf{0} \\
\mathbf{0} & \mathbf{C}_2^{[i]} & \cdots & \mathbf{0} \\
\vdots & \vdots &\ddots &\vdots \\
\mathbf{0} & \mathbf{0} & \cdots & \mathbf{C}_M^{[i]}\end{bmatrix}.\]
For instance, in the Supplemental Appendix we revisit our empirical application, where $\bA^{[i]}$ contains the (log) GDP per capita and the investment-to-GDP ratio ($M=2$) of an ever-liberalized economy $i$ during the pre-liberalization period, and $\bB^{[i]}$ contains the same two features of the donor economies used to match $\bA^{[i]}$. For each feature $l=1,2$, $\bC_l^{[i]}$ contains an intercept and a linear time trend ($K=2$).

We search for a vector of weights $\bw=(\bw^{[1]\prime}, \cdots, \bw^{[J_1]\prime})'\in\mathcal{W}\subseteq\mathbb{R}^{J_0J_1}$, which is common across the $M$ features, and a vector of coefficients $\br=(\br^{[1]\prime}, \cdots, \br^{[J_1]\prime})'\in\mathcal{R}\subseteq\mathbb{R}^{KMJ_1}$, such that the linear combination of $\bB^{[i]}$ and $\bC^{[i]}$ matches $\bA^{[i]}$ as closely as possible, for all $i\in\mathcal{E}$. The feasibility sets $\mathcal{W}$ and $\mathcal{R}$ capture the restrictions imposed.
Analogously to \eqref{eq: weights in basic framework},
such SC weights are obtained via the following optimization problem: 
for some $(\tilde{T}\cdot M)\times (\tilde{T}\cdot M)$ symmetric weighting matrix $\bV$ with $\tilde{T}=\sum_{i=1}^{J_1} T_{0i}$,
\begin{equation}\label{eq: estimated weight, general}
\widehat{\bbeta} := (\widehat{\bw}', \;\widehat{\br}')' \in\underset{\bw\in\mathcal{W},\, \br\in\mathcal{R}}{\argmin}\;
(\bA-\bB\bw-\bC\br)'\bV(\bA-\bB\bw-\bC\br)
\end{equation}
where 
\begin{footnotesize}
\[
\underbrace{\mathbf{A}}_{\tilde{T}\cdot M\times 1} = \begin{bmatrix} \mathbf{A}^{[1]} \\ \vdots \\ \mathbf{A}^{[J_1]} \end{bmatrix}, \; 
\underbrace{\mathbf{B}}_{\tilde{T}\cdot M \times J_0\cdot J_1} = \begin{bmatrix} \mathbf{B}^{[1]} & \mathbf{0} & \cdots & \mathbf{0} \\
\mathbf{0} & \mathbf{B}^{[2]} & \cdots & \mathbf{0} \\
\vdots & \vdots &\ddots &\vdots \\
\mathbf{0} & \mathbf{0} & \cdots & \mathbf{B}^{[J_1]} \end{bmatrix},\; 
\underbrace{\mathbf{C}}_{\tilde{T}\cdot M \times K\cdot M\cdot J_1}=\begin{bmatrix} \mathbf{C}^{[1]} & \mathbf{0} & \cdots & \mathbf{0} \\
\mathbf{0} & \mathbf{C}^{[2]} & \cdots & \mathbf{0} \\
\vdots & \vdots &\ddots &\vdots \\
\mathbf{0} & \mathbf{0} & \cdots & \mathbf{C}^{[J_1]}\end{bmatrix}.
\]
\end{footnotesize}

\noindent Accordingly, we write $\widehat{\bw}=(\widehat\bw^{[1]\prime}, \cdots, \widehat{\bw}^{[J_1]\prime})'$ where each $\widehat\bw^{[i]}=(\widehat{w}_1^{[i]}, \cdots, \widehat{w}_{J_0}^{[i]})'$ is the SC weights on $J_0$ donor units that are used to predict the counterfactual of the treated unit $i$. Similarly, write $\widehat\br=(\widehat{\br}^{[1]\prime}, \cdots, \widehat{\br}^{[J_1]\prime})'$ and 
$\widehat{\bbeta}=(\widehat{\bbeta}^{[1]\prime}, \cdots, \widehat\bbeta^{[J_1]\prime})'$.

\begin{remark}[Weighting Matrix]
As pointed out by \citet{BenMichael-Feller-Rothstein_2021_JRSSB}, with multiple treated units, the SC weights could be constructed in two ways: (i) optimizing the separate fit for each treated unit; and (ii) optimizing the pooled fit for the average of the treated units. These ideas can be accommodated by choosing a proper weighting matrix $\bV$. For example, assume $T_{i0}=T_0$ for simplicity. Taking $\bV=\bI_{T_0MJ_1}$ yields
\[
\widehat{\bbeta}=\underset{\bw\in\mathcal{W},\br\in\mathcal{R}}{\arg\min}\;
\sum_{i=1}^{J_1}\sum_{l=1}^M\sum_{t=1}^{T_0}
\Big(a_{t,l}^{[i]}-\bb_{t,l}^{[i]\prime}\bw_l^{[i]}-\bc_{t,l}^{[i]\prime}\br_{l}^{[i]}\Big)^2,
\]
where 
$\bB_{l}^{[i]}:=(\bb_{1,l}^{[i]}, \cdots, \bb_{T_0,l}^{[i]})'$ is the $l$-th feature of the $J_0$ donor units, and
$\bC_{l}^{[i]}:=(\bc_{1,l}^{[i]}, \cdots, \bc_{T_0,l}^{[i]})'$ is the additional $K$ variables used to predict $\bA_l^{[i]}$. 
The objective above is equivalent to minimizing the sum of squared errors of the pre-treatment fit for \textit{each} treated unit and thus is termed ``separate fit''. 
By contrast, consider the following weighting matrix:
$\bV=\frac{1}{J_1^2}\bm{1}_{J_1}\bm{1}_{J_1}'\otimes \bI_{T_0M}$
where $\otimes$ denotes the Kronecker product operator. Then, 
\[
\widehat{\bbeta}=\underset{\bw\in\mathcal{W},\br\in\mathcal{R}}{\arg\min}\;
\sum_{l=1}^M\sum_{t=1}^{T_0}\bigg[\frac{1}{J_1}\sum_{i=1}^{J_1}
\Big(a_{t,l}^{[i]}-\bb_{t,l}^{[i]\prime}\bw^{[i]}-
\bc_{t,l}^{[i]\prime}\br_{l}^{[i]}\Big)\bigg]^2.
\]
In this case, the goal is to minimize the sum of squared \textit{averaged} errors across all treated units, which is usually termed ``pooled fit''.
\end{remark}

Given the SC weights, the predicted counterfactual outcome of each treated unit $i\in\mathcal{E}$ is
\[
\widehat{Y}_{i(T_i+k)}(\infty):=
\bx_{T_i+k}^{[i]\prime}\widehat\bw^{[i]}+\bg_{T_i+k}^{[i]\prime}\widehat\br^{[i]}=\bp_{T_i+k}^{[i]\prime}\widehat{\bbeta}^{[i]},\quad
\bp_{T_i+k}^{[i]}=(\bx_{T_i+k}^{[i]\prime}, \;\bg_{T_i+k}^{[i]\prime})',
\quad k\geq 0,
\]
where $\bx_{T_i+k}^{[i]}$ is a vector of predictors of the donor units used to predict the counterfactual of the treated unit $i$ measured $k$ periods after the treatment, and $\bg_{T_i+k}^{[i]}$ is a vector of predictors that correspond to the additional control variables specified in $\bC^{[i]}$. In general, the variables included in $\bx_{T_i+k}^{[i]}$ and $\bg_{T_i+k}^{[i]}$ need not be the same as those in $\bB^{[i]}$ and $\bC^{[i]}$. 

Again, let $\tau$ denote any of the four causal predictands defined in Section \ref{sec: basic framework}.
Then, the prediction of $\tau$ can be constructed accordingly and uniformly expressed  as
\begin{equation}\label{eq: prediction of tau in general}
\widehat{\tau}=\mathsf{L}(\{Y_{it}\})-\bp_\tau'\widehat{\bbeta},
\end{equation}
with $\mathsf{L}(\{Y_{it}\})$ some linear combination of observed post-treatment outcomes and $\bp_\tau'\widehat{\bbeta}$ the prediction of the corresponding counterfactual. 
$\bp_\tau$ here denotes a predictor vector associated with the predictand $\tau$, whose specific expression in each case is as follows:
\begin{align*}
    &\bp_{\tau_{ik}}=(\underbrace{\mathbf{0}_{J_0+KM}', \cdots, \mathbf{0}_{J_0+KM}'}_{(i-1) \text{  vectors}}, 
\; \bp_{T_i+k}^{[i]\prime},\; \underbrace{\mathbf{0}_{J_0+KM}', \cdots, \mathbf{0}_{J_0+KM}'}_{(J_1-i) \text{  vectors}})',\\
&\bp_{\tau_{i \cdot}}=\Big(\underbrace{\mathbf{0}_{J_0+KM}', \cdots, \mathbf{0}_{J_0+KM}'}_{(i-1) \text{  vectors}},\; \frac{1}{T-T_i+1}\sum_{t\geq T_i}\bp_t^{[i]\prime}, \;\underbrace{\mathbf{0}_{J_0+KM}', \cdots, \mathbf{0}_{J_0+KM}'}_{(J_1-i) \text{  vectors}}\Big)',\\
&\bp_{\tau_{\mathcal{Q}k}}=\Big(
\frac{1}{Q}\bp_{T_1+k}^{[1]\prime}\I(T_1\in\mathcal{Q}),
\frac{1}{Q}\bp_{T_2+k}^{[2]\prime}\I(T_2\in\mathcal{Q}),\cdots, 
\frac{1}{Q}\bp_{T_{J_1}+k}^{[J_1]\prime}\I(T_{J_1}\in\mathcal{Q})\Big)',\quad\text{and}\\
&\bp_{\tau_{\cdot\cdot}}=\Big(\frac{1}{LJ_1}\sum_{k=1}^L\bp_{T_1+k}^{[1]\prime},
\frac{1}{LJ_1}\sum_{k=1}^L\bp_{T_2+k}^{[2]\prime},\cdots, \frac{1}{LJ_1}\sum_{k=1}^L\bp_{T_{J_1}+k}^{[J_1]\prime}\Big)',
\end{align*}
where $\mathbf{0}_{J_0+KM}$ denotes a $(J_0+KM)$-dimensional vector of zeros.

Analogously to \eqref{eq: pseudo true value in basic framework}, the pseudo-true value of SC weights in this framework is defined by
\begin{equation}\label{eq: pseudo true value, general}
\bbeta_0 := (\bw_0', \,\br_0')' = \underset{\bw\in\mathcal{W},\, \br\in\mathcal{R}}{\arg\min}\; \E\Big[(\bA-\bB\bw-\bC\br)'\bV(\bA-\bB\bw-\bC\br)\Big|\mathscr{H}\Big],
\end{equation}
and then we can write
\begin{equation}\label{eq: vertical regression}
\bA=\bB\bw_0+\bC\br_0+\bU, \qquad \bw_0\in\mathcal{W}, \qquad \br_0\in\mathcal{R},
\end{equation}
where $\bU=(\bu^{[1]\prime}, \cdots, \bu^{[J_1]\prime})'\in\mathbb{R}^{\tilde{T}M}$ is the corresponding pseudo-true residual relative to the $\sigma$-field $\mathscr{H}=\{\bB, \bC, \bp_\tau\}$.

As before, we differentiate the contribution of the in-sample error $\mathsf{InErr}(\tau)$ and the out-of-sample error $\mathsf{OutErr}(\tau)$ to the uncertainty of SC prediction of $\tau$. For the in-sample uncertainty, the optimization bounds used in Section \ref{sec: in-sample error} can be generalized to this setup. 
Specifically, let $d_\beta=J_0+KM$, $\bZ=(\bB, \bC)$, $\widehat\bQ=\bZ'\bV\bZ$, $\widehat\bgamma'=\bU'\bV\bZ$,
$\bgamma=\E[\widehat{\bgamma}|\mathscr{H}]$,
and $\Delta=\{\bdelta\in\mathbb{R}^{d_\beta}: \bdelta+\bbeta_0\in\mathcal{W}\times \mathcal{R}\}$.
It follows from the optimality of $\widehat{\bbeta}$ and $\bbeta_0$ and the convexity of $\mathcal{W}$ and $\mathcal{R}$ that  
$\widehat{\bbeta}-\bbeta_0
\in\Delta$ and 
$(\widehat{\bbeta}-\bbeta_0)'\widehat{\bQ}(\widehat{\bbeta}-\bbeta_0)-
2(\widehat{\bgamma}-\bgamma)'(\widehat{\bbeta}-\bbeta_0)\leq 0$.
Also, given the expression \eqref{eq: prediction of tau in general}, $\mathsf{InErr}(\tau)$ can be generally expressed as $-\bp_\tau(\widehat{\bbeta}-\bbeta_0)$. 
Then, a valid, though infeasible, prediction interval for the in-sample error $\mathsf{InErr}(\tau)$ is 
$[\underline{\mathfrak{c}}(\alpha_{\tin}/2),\;\overline{\mathfrak{c}}(1-\alpha_{\tin}/2)]$, 
where
$\underline{\mathfrak{c}}(\alpha)$ denotes the $\alpha$-quantile of $\inf_{\bdelta\in\mathcal{M}_{\bG}}-\bp_{\tau}'\bdelta$
and $\overline{\mathfrak{c}}(\alpha)$ denotes $\alpha$-quantile of $\sup_{\bdelta\in\mathcal{M}_{\bG}}
-\bp_{\tau}'\bdelta$,
conditional on $\mathscr{H}$, for any $\alpha\in(0,1)$, with $\mathcal{M}_{\bG}=\{\bdelta\in\Delta:\bdelta'\widehat{\bQ}\bdelta-2\bG'\bdelta\leq 0\}$, $\bG|\mathscr{H}\thicksim\mathsf{N}(\mathbf{0}, \bSigma)$ and  $\bSigma=\V[\widehat{\bgamma}|\mathscr{H}]$.
Analogously to \eqref{eq: in-sample error bound, basic setup}, once a feasible variance estimator $\widehat{\bSigma}$ of $\bSigma$ and a feasible constraint set $\Delta^\star$ that contains the original $\Delta$ in the small neighborhood of zero (as described by condition (iii) in Theorem \ref{thm: coverage error approximation, plug-in} below) are available, we can set
\begin{equation}\label{eq: bound on in-sample error}
\myuline{M}_{\tin}(\tau)=\underline{\mathfrak{c}}^\star(\alpha_{\tin}/2)
\quad\text{and}\quad
\myoline{M}_{\tin}(\tau)=\bar{\mathfrak{c}}^\star(1-\alpha_{\tin}/2)
\end{equation}
where 
$\underline{\mathfrak{c}}^\star(\alpha_{\tin}/2)$ is the $(\alpha_{\tin}/2)$-quantile of $\inf_{\bdelta\in\mathcal{M}_{\bG}^\star}
-\bp_{\tau}'\bdelta$, and 
$\bar{\mathfrak{c}}^\star(1-\alpha_{\tin}/2)$ is the $(1-\alpha_{\tin}/2)$-quantile of $\sup_{\bdelta\in\mathcal{M}_{\bG}^\star}
-\bp_{\tau}'\bdelta$ conditional on the data, 
with $\mathcal{M}_{\bG}^\star=\{\bdelta\in\Delta^\star:\;
\bdelta'\widehat\bQ\bdelta-2(\bG^\star)'\bdelta\leq 0\}$ and 
$\bG^\star|\mathsf{Data}\sim\mathsf{N}(\bm{0},\widehat\bSigma)$.

For the out-of-sample error, analogously to \eqref{eq: out-of-sample error bound, basic setup}, if $\mathsf{OutErr}(\tau)$ is assumed to be sub-Gaussian conditional on $\mathscr{H}$ with parameter $\sigma_{\mathscr{H}}$, then we can set
\begin{equation}\label{eq: bound on out-of-sample error}
\begin{aligned}	
    \myuline{M}_{\tout}(\tau)&=\E[\mathsf{OutErr}(\tau)|\mathscr{H}]-\sqrt{2\sigma_{\mathscr{H}}^2\log(2/\alpha_{\tout})} 
	\quad\text{and}\\
	\myoline{M}_{\tout}(\tau)&=\E[\mathsf{OutErr}(\tau)|\mathscr{H}]+\sqrt{2\sigma_{\mathscr{H}}^2\log(2/\alpha_{\tout})}.
\end{aligned}	
\end{equation}

We present a general theorem that justifies the above method under high-level conditions and covers the results given in Section \ref{sec: prediction interval} as a special case. Let $\|\cdot\|_\mathtt{F}$ be the Frobenius matrix norm, $\|\cdot\|$ any $\ell_p$ vector norm with $p\geq 1$, and $\mathcal{B}(\mathbf{0}, c)=\{\blambda\in\mathbb{R}^{d_\beta}:\|\blambda\|\leq c\}$ a $c$-neighborhood of zero for any $c>0$.

\begin{thm}
	\label{thm: coverage error approximation, plug-in}
	Assume $\mathcal{W}$ and $\mathcal{R}$ are convex, $\widehat{\bbeta}$ in \eqref{eq: estimated weight, general} and $\bbeta_0$ in \eqref{eq: pseudo true value, general} exist,  $\mathscr{H} = \sigma(\bB, \bC, \bp_{\tau})$, and
	$\myuline{M}_{\tin}(\tau)$, $\myoline{M}_{\tin}(\tau)$, $\myuline{M}_{\tout}(\tau)$ and
	$\myoline{M}_{\tout}(\tau)$ are specified as in \eqref{eq: bound on in-sample error} and \eqref{eq: bound on out-of-sample error}. In addition, for some finite non-negative constants $\epsilon_{\gamma}$, $\pi_{\gamma}$, $\varpi_\delta^{\ttb}$, $\epsilon_{\delta}^{\ttb}$, $\pi_\delta^{\ttb}$, $\epsilon_{\Delta}^{\ttb}$, $\pi_{\Delta}^{\ttb}$, $\epsilon_{\gamma,1}^{\ttb}$, $\epsilon_{\gamma,2}^{\ttb}$ and  $\pi_{\gamma}^{\ttb}$, the following conditions hold:
	\begin{enumerate}[label=\normalfont(\roman*),noitemsep]
		\item $\P[\P(-\bp_{\tau}'(\widehat{\bbeta}-\bbeta_0)\in [\underline{\mathfrak{c}}(\alpha_0), \overline{\mathfrak{c}}(1-\alpha_0)]|\mathscr{H})\geq 1-2\alpha_0-\epsilon_\gamma]\geq 1-\pi_\gamma$ for any $\alpha_0\in(0,1)$;
		\item $\P[\P(\sup\{\|\bdelta\|:	\bdelta\in\mathcal{M}_\bG\} \leq \varpi_\delta^{\ttb}|\mathscr{H})\geq 1-\epsilon_{\delta}^{\ttb}]\geq 1-\pi_\delta^{\ttb}$;
	\item $\P[\P(\Delta\cap\mathcal{B}(\bm{0},\varpi_\delta^{\ttb})\subseteq
		\Delta^\star \;|\mathscr{H})\geq 1-\epsilon_{\Delta}^{\ttb}]\geq 1-\pi_{\Delta}^{\ttb}$;
		\item $\P[\P(\|\bSigma^{-1/2}\widehat{\bSigma}\bSigma^{-1/2}-\bI_{d_\beta}\|_\mathtt{F}\leq 2\epsilon_{\gamma,1}^{\ttb}|\mathscr{H})\geq 1-\epsilon_{\gamma,2}^{\ttb}]\geq 1-\pi_{\gamma}^{\ttb}$;
		\item $\mathsf{OutErr}(\tau)-\E[\mathsf{OutErr}(\tau)|\mathscr{H}]$ is sub-Gaussian conditional on $\mathscr{H}$ with parameter $\sigma_{\mathscr{H}}$.
	\end{enumerate}
    Then, for $\epsilon_{\gamma,1}^{\ttb}\in[0,1/4]$,
	\[
	\P\Big\{\P\big(\tau\in[
	\widehat{\tau}-\myoline{M}_{\tin}(\tau)-\myoline{M}_{\tout}(\tau),\;
	\widehat{\tau}-\myuline{M}_{\tin}(\tau)-\myuline{M}_{\tout}(\tau)]
    \big|\mathscr{H}\big)\geq 1-\alpha_{\tin}-\alpha_{\tout}-\epsilon\Big\}\geq 1-\pi,
	\]
	where $\epsilon=\epsilon_{\gamma}+2\epsilon_{\gamma,1}^{\ttb}+\epsilon_{\gamma,2}^{\ttb}+2\epsilon_{\delta}^{\ttb}+\epsilon_{\Delta}^{\ttb}$ and 
    $\pi=\pi_{\gamma}+\pi_\gamma^{\ttb}+\pi_{\delta}^{\ttb}+\pi_{\Delta}^{\ttb}$.
\end{thm}

Assumptions (i)--(iv) imposed in Theorem \ref{thm: coverage error approximation, plug-in} are high-level, which are used for in-sample uncertainty quantification and can be verified in many practically relevant scenarios such as the cointegrated data considered in Section \ref{sec: prediction interval}. We give more detailed discussion of each condition in Section \ref{subsec: high-level conditions} below. 
Assumption (v), as we emphasized before, is a moment condition used to showcase our out-of-sample uncertainty quantification strategy and can be relaxed by utilizing other appropriate concentration inequalities.

\subsection{Discussion of Conditions (i)--(iv)}\label{subsec: high-level conditions}

In this section, we discuss the justification of the high-level conditions (i)--(iv) in more detail.

\begin{itemize}[leftmargin=*]
    \item \textbf{Condition (i)}. This condition formalizes the idea of distributional approximation of $\widehat{\bgamma}-\bgamma$ by a Gaussian vector $\bG$. It can be verified under different primitive conditions, such as Assumption \ref{assumption: dgp} that accommodates non-stationary data and is applicable to our empirical application.
    Lemma S.1 in the Supplemental Appendix provides a general way to verify condition (i) by assuming the the pseudo-true residuals in $\bU$ are independent over $t$ conditional on $\mathscr{H}$. In fact, (i) also holds when the errors are only weakly dependent (e.g., $\beta$-mixing) conditional on $\mathscr{H}$.
    
    \item \textbf{Condition (ii)}. This is a mild condition on the concentration of $\bdelta\in \mathcal{M}_{\bG}$. The requirement  $\bdelta'\widehat{\bQ}\bdelta-2\bG'\bdelta\leq 0$ is usually known as the \textit{basic inequality} in regression analysis; see, for example, \citet[Chapter 7]{Wainwright_2019_Book} for the case of lasso. The vector $\bG$ is (conditionally) Gaussian by construction, making condition (ii) easy to verify based on well-known bounds for Gaussian distributions. This condition holds in a variety of empirically relevant settings, including outcomes-only regression with i.i.d. data, multi-equation regression with weakly dependent data, and settings with cointegrated outcomes and features.
    
    \item \textbf{Condition (iii)}. This is a high-level requirement on the constraint set $\Delta^\star$ used in the simulation. Intuitively, to obtain valid bounds on the in-sample error, the supremum or infimum of $-\bp_\tau'\bdelta$ should be searched for over a set of $\bdelta$ values that contains $\Delta$, or at least the portion of it within a small neighborhood of zero. One simple example that always satisfies condition (iii) is $\Delta^\star=\{(\bw_1'-\bw_2', \br_1'-\br_2')': (\bw_1,\br_1), (\bw_2,\br_2)\in\mathcal{W}\times\mathcal{R}\}$. However, this set is typically large, leading to overly conservative bounds. Thus, we provide an improved, general strategy to construct $\Delta^\star$ in this setting. It can be shown to satisfy condition (iii) if the constraints specified in $\mathcal{W}$ and $\mathcal{R}$ are formed by smooth functions. Suppose that 
    $$
    \mathcal{W}\times\mathcal{R}=\Big\{\bbeta\in\mathbb{R}^{d_\beta}:\bm{m}_{=}(\bbeta)=\bm{0},\bm{m}_{\leq}(\bbeta)\leq \bm{0}\Big\},
    $$
    where $\bm{m}_{=}(\cdot)\in\mathbb{R}^{d_{=}}$ and $\bm{m}_{\leq}(\cdot)\in\mathbb{R}^{d_{\leq}}$ and $d_{=}$ and $d_{\leq}$ denote the number of equality and inequality constraints in $\mathcal{W}\times\mathcal{R},$ respectively. 
    Let the $\ell$-th constraint in $\bm{m}_{\leq}(\cdot)$ be $m_{\leq,\ell}(\cdot)$. Given tuning parameters $\varrho_\ell>0$, $\ell=1, \cdots, d_{\leq}$, let $\mathcal{A}=\{\ell_1, \cdots, \ell_k\}$ denote the set of indices for the inequality constraints such that $m_{\leq, \ell}(\widehat{\bbeta})>-\varrho_\ell$. We define
    \begin{equation}\label{eq: Delta*}
    \widehat{\Delta}=\Big\{
    \bbeta-\widehat{\bbeta}:\bm{m}_{=}(\bbeta)=\bm{0},\; 
    m_{\leq,\ell}(\bbeta)\leq 
    m_{\leq,\ell}(\widehat{\bbeta})\I(\ell\in\mathcal{A}), \; \ell=1, \cdots, d_{\leq}
    \Big\}.
    \end{equation}
    Then, let $\Delta^\star=\widehat{\Delta}_\varepsilon:=\{\bdelta: \mathrm{dist}(\bdelta,\widehat{\Delta})\leq\varepsilon\}$, where $\mathrm{dist}(\bdelta, \widehat{\Delta})=\inf_{\blambda\in\widehat{\Delta}}\|\bdelta-\blambda\|_2$. That is, $\Delta^\star$ is an ``$\varepsilon$-enlargement'' of $\widehat{\Delta}$ for some $\varepsilon\geq 0$. The following lemma verifies condition (iii) for this $\Delta^\star$. 

    \begin{lem} \label{lem: nonlinear constraints}
     Assume that with probability over $\mathscr{H}$ at least $1-\pi_\Delta^\star$, the following conditions hold: 
    (i) $\P(\|\widehat\bbeta-\bbeta_0\|_2\leq \varpi_\delta^\star|\mathscr{H})\geq 1-\epsilon_\Delta^\star$; 
    (ii) $\bm{m}(\cdot)=(\bm{m}_{=}(\cdot)',\bm{m}_{\leq}(\cdot)')'$ is twice continuously differentiable on $\mathcal{B}(\bbeta_0, \varpi_{\delta}^\star)$ with $\inf_{\bbeta\in\mathcal{B}(\bbeta_0,\varpi_\delta^\star)}s_{\min}(\frac{\partial}{\partial\bbeta'}\bm{m}(\bbeta))\geq c_{\min}$ for some constant $c_{\min}>0$;  
    and (iii) for all $1\leq \ell\leq d_{\leq}$ and some $\mathfrak{c}>0$ specified in the proof, 
    $\varrho_\ell>\mathfrak{c}\varpi_\delta^\star$, and
    $\varrho_\ell<|m_{\leq,\ell}(\bbeta_0)|-\mathfrak{c}\varpi_\delta^\star$ if
    $m_{\leq, \ell}(\bbeta_0)\neq 0$.
    Then, condition (iii) of Theorem \ref{thm: coverage error approximation, plug-in} holds for $\Delta^\star=\widehat{\Delta}_\varepsilon$ with  $\varepsilon\geq\mathfrak{C}(\varpi_{\delta}^\star)^2$ for some constant $\mathfrak{C}>0$. If $\bm{m}(\bbeta)$ is linear in $\bbeta$, condition (iii) holds for $\Delta^\star=\widehat{\Delta}$.
    \end{lem}

    The strategy described in Section \ref{sec: in-sample error} is an application of this theoretical result, as detailed in Section S.5.2.2 of the Supplemental Appendix.
    In this lemma, the tuning parameters $\varrho_\ell$'s are introduced to guarantee that, with high probability, we can correctly differentiate the binding inequality constraints from the other non-binding ones. 
    Section \ref{subsec: define constraints in simulation} below provides more practical details about choosing $\varrho_\ell$. Also, the concentration requirement for  $\widehat\bbeta$ specified in this lemma is mild. Since $\widehat\bbeta$ satisfies the basic inequality 
    $(\widehat{\bbeta}-\bbeta_0)'\widehat{\bQ}(\widehat{\bbeta}-\bbeta_0)-
    2(\widehat{\bgamma}-\bgamma)'(\widehat{\bbeta}-\bbeta_0)\leq 0$, the concentration of $\widehat\bbeta$ can be shown by combining a distributional approximation of $\widehat\bgamma-\bgamma$ by a Gaussian vector $\bG$ and the idea outlined in the previous discussion about condition (ii).

    \item \textbf{Condition (iv)}. This is a requirement that $\widehat{\bSigma}$ be a ``good'' approximation of the unknown covariance matrix $\bSigma$. Many standard covariance estimation strategies such as the family of well-known heteroskedasticity-consistent estimators can be utilized.
\end{itemize}

\subsection{Defining Constraints in Simulation}
\label{subsec: define constraints in simulation}

We propose a feasible strategy to construct the constraint set $\Delta^\star$ used in the simulation. 
We focus on the most common case in practice: each constraint only restricts the SC weights $\widehat{\bbeta}^{[i]}$ for one treated unit $i$, so there is no ``cross-treated-unit'' constraint.

First, we introduce the tuning parameters $\varrho_\ell$, $\ell=1, \cdots, d_{\leq}$, to determine which inequality constraints are binding. We define $\varrho_\ell$ as a high-probability bound on $m_{\leq,\ell}(\widehat{\bbeta}^{[i]})$. By Taylor expansion, if the constraint $m_{\leq, \ell}(\bbeta^{[i]})\leq 0$ is binding (i.e., $m_{\leq, \ell}(\bbeta_0^{[i]})=0$), then $m_{\leq,\ell}(\widehat{\bbeta}^{[i]})\approx
\frac{\partial}{\partial \bbeta'}m_{\leq,\ell}(\bbeta_0^{[i]})(\widehat\bbeta^{[i]}-\bbeta_0^{[i]})$. Therefore, given a high-probability bound $\varrho^{[i]}$ on $\|\widehat\bbeta^{[i]}-\bbeta_0^{[i]}\|_2$, we set
\begin{align}
    \varrho_\ell=\Big\|\frac{\partial}{\partial\bbeta}m_{\leq,\ell}(\widehat{\bbeta}^{[i]})\Big\|_2\times \varrho^{[i]}.
\label{eq: tuning varrho_j}
\end{align}

Next, to select $\varrho^{[i]}$, we employ the basic inequality $(\widehat\bbeta^{[i]}-\bbeta_0^{[i]})'\widehat{\bQ}(\widehat\bbeta^{[i]}-\bbeta_0^{[i]})-2(\widehat{\bgamma}-\bgamma)'(\widehat\bbeta^{[i]}-\bbeta_0^{[i]})\leq 0$. As mentioned before, it always holds by optimality of $\widehat{\bbeta}$, which, combined with the Gaussian approximation of $\widehat{\bgamma}-\bgamma$ by $\bG$, implies a high-probability deviation bound for $\widehat{\bbeta}^{[i]}$: $\|\widehat{\bbeta}-\bbeta_0\|_2\leq 2\|\bG\|_2/s_{\min}(\widehat{\bQ})$. Motivated by this fact, 
we propose to set $\varrho^{[i]} = \mathcal{C}T_0^{-1/2}$, where $\mathcal{C}$ is defined as
\begin{align}\label{eq: tuning varrho}
    \mathcal{C} = \frac{\sqrt{d_{\beta,0}\log(d_\beta) \log(T_0)}\max_{1\leq j \leq J_0}\widehat{\sigma}_{b_j}\widehat{\sigma}_{u}}{\min_{1\leq j \leq J_0}\widehat{\sigma}^2_{b_j}},
\end{align}
where $d_{\beta,0}$ denotes the number of nonzeros in $\widehat{\bbeta}$, and $\widehat\sigma_u$ and $\widehat\sigma_{b_j}$ are the estimated (unconditional) standard deviation of $\bu^{[i]}$ and the $j$-th column of $\bB^{[i]}$, respectively. If the SC weights are constructed by matching on both stationary and non-stationary features, the precision of the estimation is primarily determined by the non-stationary components. In such cases, one may disregard the stationary components when determining $\mathcal{C}$ using \eqref{eq: tuning varrho}. See Supplemental Appendix Section S.5.1 for more explanation. Moreover, the constant in \eqref{eq: tuning varrho} is a ``rule-of-thumb'' choice that can be rationalized under specific conditions and at least have the correct order of magnitude for $\|\widehat{\bbeta}-\bbeta_0\|_2$. Details are provided in Section S.5.1 of the Supplemental Appendix. 

Then, in the simulation, we impose
\begin{equation}\label{eq: enlargement}
m_{\leq,\ell}(\bbeta^{[i]})\leq m_{\leq,\ell}(\widehat{\bbeta}^{[i]})+\frac{1}{2}s_{\max}\Big(\frac{\partial}{\partial\bbeta\partial\bbeta'}m_{\leq,\ell}(\widehat{\bbeta}^{[i]})\Big)\times
(\varrho^{[i]})^2
\end{equation}
if $m_{\leq,\ell}(\widehat{\bbeta}^{[i]})>-\varrho_\ell$ (``binding''), and retain $m_{\leq, \ell}(\bbeta^{[i]})\leq 0$ otherwise (``non-binding''). This proposed adjusted constraint is motivated by the characterization of the distance between $\Delta$ and $\widehat{\Delta}$ in Lemma \ref{lem: nonlinear constraints}, which typically depends on the second-order expansion of the constraint function around the true values. See Section S.5.2.2 of the Supplemental Appendix for more discussion. If $m_{\leq,\ell}(\cdot)$ is linear, the second-order derivative of $m_{\leq,\ell}(\cdot)$ is exactly zero, leading to the constraint $m_{\leq,\ell}(\bbeta^{[i]})\leq m_{\leq, \ell}(\widehat{\bbeta}^{[i]})$. Thus, the adjustment due to the second term on the right-hand side of \eqref{eq: enlargement} is only necessary for nonlinear constraints. This coincides with the result in Lemma \ref{lem: nonlinear constraints}. In the special case of the L2 constraint used in our empirical application, \eqref{eq: enlargement} implies that we impose $\|\bbeta^{[i]}\|_2^2\leq \|\widehat{\bbeta}^{[i]}\|_2^2+(\varrho^{[i]})^2$ in the simulation if $m_{\leq,\ell}(\bbeta^{[i]})\leq 0$ is determined to be a binding constraint. 

\section{Conclusion}

We developed prediction intervals to quantify the uncertainty of a large class of synthetic control predictions (or estimators) in settings with staggered treatment adoption. Because many synthetic control applications have a limited number of observations, our inference procedures are based on non-asymptotic concentration arguments. The construction of our prediction intervals is designed to capture two sources of uncertainty: the first is the construction or estimation of the synthetic control weights with pre-treatment data, and the second is the variability of the post-treatment outcomes. By combining both sources in a prediction interval, our procedure offers precise non-asymptotic coverage probability guarantees and allows researchers to implement sensitivity analyses to assess how robust the conclusions of the analysis are to various levels of uncertainty. Our framework is general, allowing for one or multiple treated units, simultaneous or staggered treatment adoption, linear or non-linear constraints, and stationary or non-stationary data. To enhance implementation, we also showed how to recast the methods as conic optimization programs and how to choose the necessary tuning parameters in a principled data-driven way. We illustrated our methods with an empirical application studying the effect of economic liberalization on real GDP per capita in Sub-Saharan African countries, motivated by the work of \cite{Billmeier-Nannicini_2013_RESTAT}.

All our methods are implemented in \texttt{Python}, \texttt{R}, and \texttt{Stata} software, which is publicly available (\url{https://nppackages.github.io/scpi/}), and discussed in detail in our companion article \citet*{Cattaneo-Feng-Palomba-Titiunik_2025_JSS} and in Section S.7 of the Supplemental Appendix.

\bibliography{CFPT_2025_ReStat--bib}
\bibliographystyle{jasa}


\end{document}